\documentclass[twocolumn,trackchanges]{aastex6}
\usepackage{multirow}
\newcommand{\Oh}{\mathcal{O}}

\begin{document}

\title{A SEARCH FOR FAST RADIO BURSTS WITH THE GBNCC PULSAR SURVEY}
\author{P. Chawla\altaffilmark{1}, V. M. Kaspi\altaffilmark{1}, A. Josephy\altaffilmark{1}, K. M. Rajwade\altaffilmark{2,3}, D. R. Lorimer\altaffilmark{2,3,4}, A. M. Archibald\altaffilmark{5}, M. E. DeCesar\altaffilmark{6}, J. W. T. Hessels\altaffilmark{5,7}, D. L. Kaplan\altaffilmark{8}, C. Karako-Argaman\altaffilmark{1}, V. I. Kondratiev\altaffilmark{5,9}, L. Levin\altaffilmark{10}, R. S. Lynch\altaffilmark{4},  M. A. McLaughlin\altaffilmark{2,3}, S. M. Ransom\altaffilmark{11}, M. S. E. Roberts\altaffilmark{12}, I. H. Stairs\altaffilmark{13}, K. Stovall\altaffilmark{14}, J. K.  Swiggum\altaffilmark{8}} \and \author{J. van Leeuwen\altaffilmark{5,7}}

\altaffiltext{1}{Department of Physics \& McGill Space Institute, McGill University, 3600 University Street, Montreal, QC H3A 2T8, Canada; pragya.chawla@mail.mcgill.ca}
\altaffiltext{2}{Department of Physics and Astronomy, West Virginia University, Morgantown, WV 26506, USA} 
\altaffiltext{3}{Center for Gravitational Waves and Cosmology, West Virginia University, Chestnut Ridge Research Building, Morgantown, WV 26505, USA}
\altaffiltext{4}{Green Bank Observatory, PO Box 2, Green Bank, WV, 24944, USA}
\altaffiltext{5}{ASTRON, the Netherlands Institute for Radio Astronomy, Postbus 2, 7990 AA Dwingeloo, The Netherlands}
\altaffiltext{6}{Department of Physics, Lafayette College, Easton, PA 18042, USA}
\altaffiltext{7}{Anton Pannekoek Institute for Astronomy, University of Amsterdam, Science Park 904, 1098 XH Amsterdam, The Netherlands}
\altaffiltext{8}{Department of Physics, University of Wisconsin-Milwaukee, Milwaukee, WI 53211, USA}
\altaffiltext{9}{Astro Space Center, Lebedev Physical Institute, Russian Academy of Sciences, Profsoyuznaya str. 84/32, 117997 Moscow, Russia}
\altaffiltext{10}{Jodrell Bank Centre for Astrophysics, School of Physics and Astronomy, The University of Manchester, Manchester, M13 9PL, UK}
\altaffiltext{11}{National Radio Astronomy Observatory, 520 Edgemont Road, Charlottesville, VA 22901, USA}
\altaffiltext{12}{New York University, Abu Dhabi, U.A.E.}
\altaffiltext{13}{Department of Physics and Astronomy, University of British Columbia, 6224 Agricultural Road, Vancouver, BC V6T 1Z1, Canada}
\altaffiltext{14}{National Radio Astronomy Observatory, 1003 Lopezville Rd., Socorro, NM 87801, USA} 

\begin{abstract}

We report on a search for Fast Radio Bursts (FRBs) with the Green Bank Northern Celestial Cap (GBNCC) Pulsar Survey at 350 MHz. Pointings amounting to a total on-sky time of 61 days were searched to a DM of 3000 pc cm$^{-3}$ while the rest (23 days; 29\% of the total time) were searched to a DM of 500 pc cm$^{-3}$. No FRBs were detected in the pointings observed through May 2016. We estimate a 95\% confidence upper limit on the FRB rate of 3.6$\times 10^3$ FRBs sky$^{-1}$ day$^{-1}$ above a peak flux density of 0.63 Jy at 350 MHz for an intrinsic pulse width of 5 ms. We place constraints on the spectral index $\alpha$ by running simulations for different astrophysical scenarios and cumulative flux density distributions. The non-detection with GBNCC is consistent with the 1.4-GHz rate reported for the Parkes surveys for $\alpha > +0.35 $ in the absence of scattering and free-free absorption and $\alpha > -0.3$ in the presence of scattering, for a Euclidean flux distribution. The constraints imply that FRBs exhibit either a flat spectrum or a spectral turnover at frequencies above 400 MHz. These constraints also allow estimation of the number of bursts that can be detected with current and upcoming surveys. We predict that CHIME may detect anywhere from several to $\sim$50 FRBs a day (depending on model assumptions), making it well suited for interesting constraints on spectral index, the log $N$-log $S$ slope and pulse profile evolution across its bandwidth (400--800 MHz).
￼
\end{abstract}

\section{Introduction} \label{sec:intro} 
Fast Radio Bursts (FRBs) are bright, millisecond-duration events occurring in the radio sky. Their origin is still unknown. Eighteen FRBs have been detected within the past decade (\citealt{lorimer2007}; \citealt{keane2012}; \citealt{thornton2013}; \citealt{BSB2014}; \citealt{masui2015}; \citealt{petroff2015}; \citealt{ravi2015}; \citealt{champion2016}; \citealt{keane2016}; \citealt{ravi2016}) with only one source (\citealt{spitler2014}; \citealt{spitler2016}) known to repeat. A catalog of these bursts and their properties is made available by \citet{petroff2016}\footnote{\url{http://www.astronomy.swin.edu.au/pulsar/frbcat/}}. These transient events can be distinguished from pulsars and rotating radio transients (RRATs) on the basis of their dispersion measure (DM), which is a measure of the integrated free electron density along the line of sight in the intervening medium. The bursts have DMs that are 1.4 to 35 times the maximum predicted along the line of sight by the NE2001 model of electron density in our Galaxy \citep{cordes2002}. 

The dominant contribution to the excess DM of FRBs can arise from the intergalactic medium, the host galaxy of the FRB progenitor, or possibly from a high electron density, compact structure in our Galaxy. The interferometric localization of bursts from the repeating FRB121102 provides evidence of its association with an optical counterpart \citep{chatterjee2017}. Spectroscopic follow-up by \citet{tendulkar2017} confirms the optical counterpart as being the host galaxy of the FRB and characterizes it as a low-metallicity, star-forming dwarf galaxy located at a redshift of $z$ = 0.19273(8). The observations of \citet{masui2015} also support an extragalactic origin with scattering and scintillation in FRB110523 suggesting that the majority of the scattering originates from within the typical size scale of a galaxy. These observations lend support to models with extragalactic progenitors of FRBs such as giant pulses from extragalactic neutron stars \citep{cordes2016b} and magnetar giant flares (\citealt{popov2013}; \citealt{kulkarni2015}). Interferometric localizations of more FRBs are essential to conclusively determine the source of the excess DM and the nature of the FRB progenitors for the broader FRB population.

All but one known FRB \citep{masui2015} has been detected at frequencies greater than 1 GHz. Detection or stringent limits at lower frequencies are crucial for understanding properties of FRBs such as their spectral index and pulse profile evolution with frequency. Searches at low frequencies with telescopes such as LOFAR (\citealt{coenen2014}; \citealt{karastergiou2015}), Arecibo \citep{deneva2016} and MWA (\citealt{tingay2015}; \citealt{rowlinson2016}) have so far not resulted in any detections. \citet{deneva2016} report an upper limit on the FRB rate at 327 MHz of 10$^{5}$ FRBs sky$^{-1}$ day$^{-1}$ for a flux density threshold of 83 mJy and pulse width of 10 ms. A non-detection with the LOFAR Pilot Pulsar Survey at 142 MHz allowed \citet{coenen2014} to place an upper limit of 150 FRBs sky$^{-1}$ day$^{-1}$, for bursts brighter than 107 Jy at burst duration 0.66 ms. \citet{karastergiou2015} report an upper limit of 29 FRBs sky$^{-1}$ day$^{-1}$ for bursts with flux density above 62 Jy at 145 MHz and a pulse width of 5 ms, based on observations with the UK station of the LOFAR radio telescope. The upper limits on the FRB rate reported thus far from these low-frequency radio surveys are not particularly constraining because of limitations in total observing time and volume searched. With observations to date amounting to a total on-sky time of 84 days, the Green Bank Northern Celestial Cap (GBNCC) Pulsar Survey \citep{stovall2014} can provide the strongest constraints yet on the FRB rate and spectral index in the frequency range of 300--400 MHz.

The GBNCC survey is also important for predicting the FRB yield of upcoming low-frequency telescopes such as the Canadian Hydrogen Intensity Mapping Experiment (CHIME). With its large field of view and good sensitivity, CHIME is predicted to discover tens of FRBs per day (\citealt{connor2016}; \citealt{rajwade2016}) in its frequency range of 400--800 MHz. The GBNCC survey is thus well placed to determine the expected detection rate for the lower part of the CHIME band. 

In this paper, we present results from the search for FRBs in GBNCC survey pointings observed through May 2016. For the purpose of our search and subsequent analysis, we define an FRB as an astrophysical pulse with a DM greater than twice the maximum line-of-sight Galactic DM. The suggestion by \citet{bannister2014} of a possibly Galactic origin of the excess DM of the only FRB with a DM ratio $<$ 2, FRB010621 \citep{keane2012}, lends support to our choice of a DM ratio of 2 for the FRB definition. 

Our paper is organized as follows. In Section \ref{sec:obser}, we give a description of the survey and its sensitivity. We describe the data analysis pipeline in Section \ref{sec:analysis} and place constraints on the FRB rate in Section \ref{sec:rate}. In Section \ref{sec:spectra}, we constrain the mean spectral index of FRBs by performing Monte-Carlo simulations of a population of FRBs. We discuss the implications for current and upcoming surveys in Section \ref{sec:implication} and present our summary and conclusions in Section \ref{sec:summary}. 

\section{Observations} \label{sec:obser}

\subsection{Survey Description}

\begin{figure*}[ht!]
\gridline{\fig{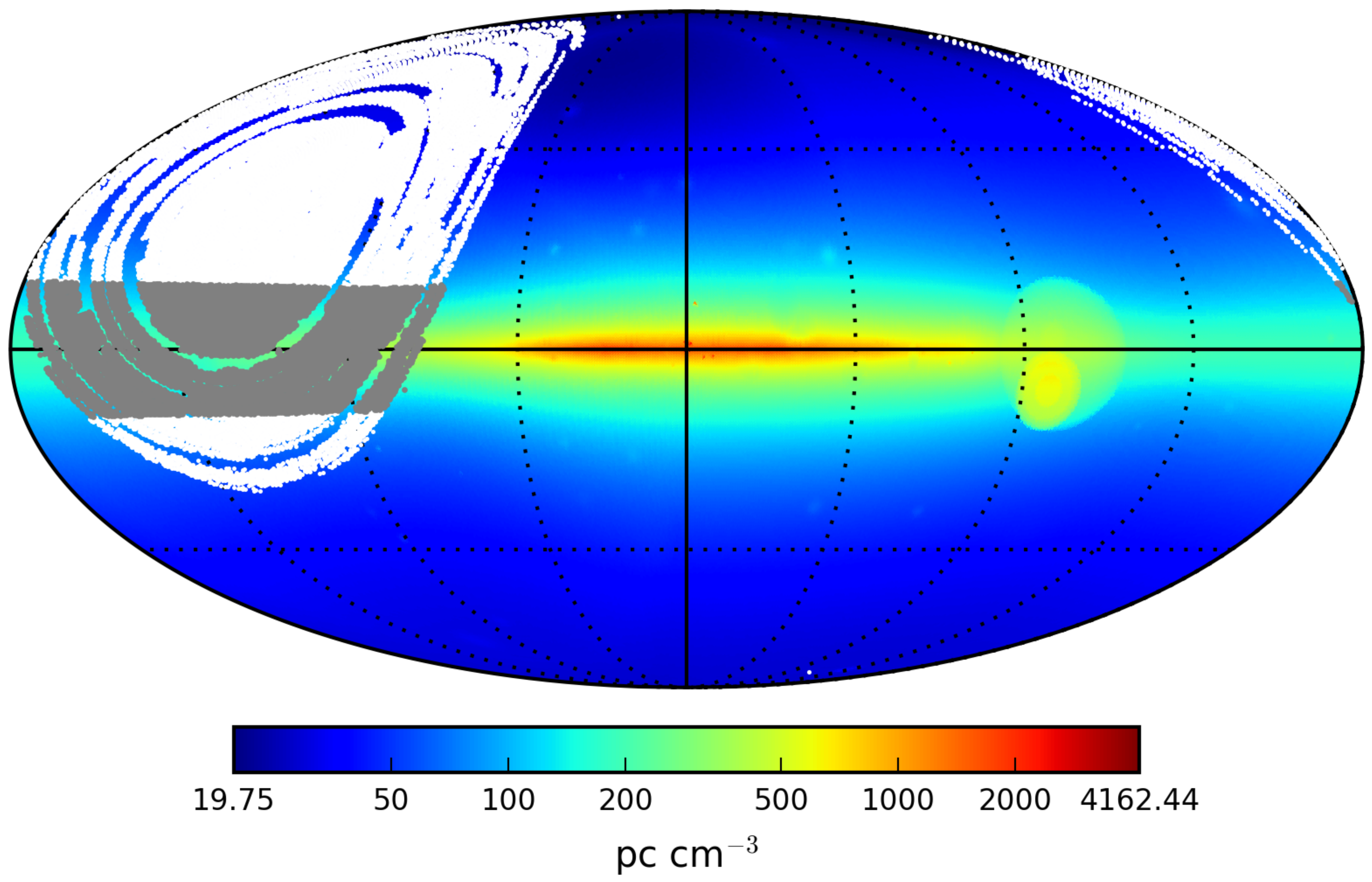}{0.43\textwidth}{(a)}
          \fig{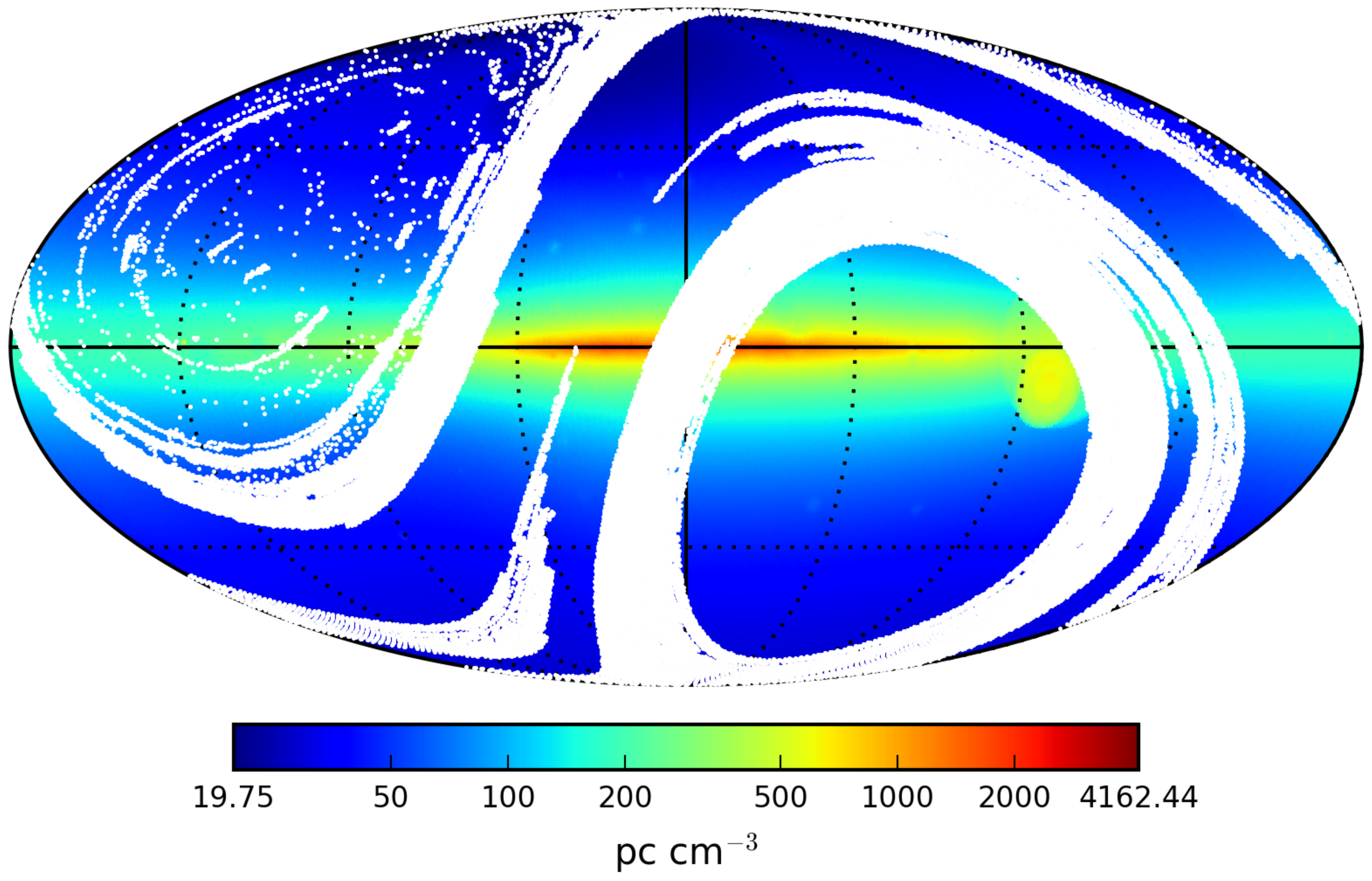}{0.43\textwidth}{(b)}
          }
\caption{Full sky map in Galactic coordinates with GBNCC pointings (marked in white) overlaid on the maximum Galactic DM predicted by the NE2001 model. Panel (a) shows the pointings searched to a DM of 500 pc cm$^{-3}$ with the excluded pointings having a predicted maximum DM $>$100 pc cm$^{-3}$ marked in grey and panel (b) shows the pointings searched to a DM of 3000 pc cm$^{-3}$. Pointings rendered unusable by the presence of RFI have not been plotted here.}
\label{fig:skymap}
\end{figure*}

\begin{figure}[hb!]
\centering
\includegraphics[scale=0.55]{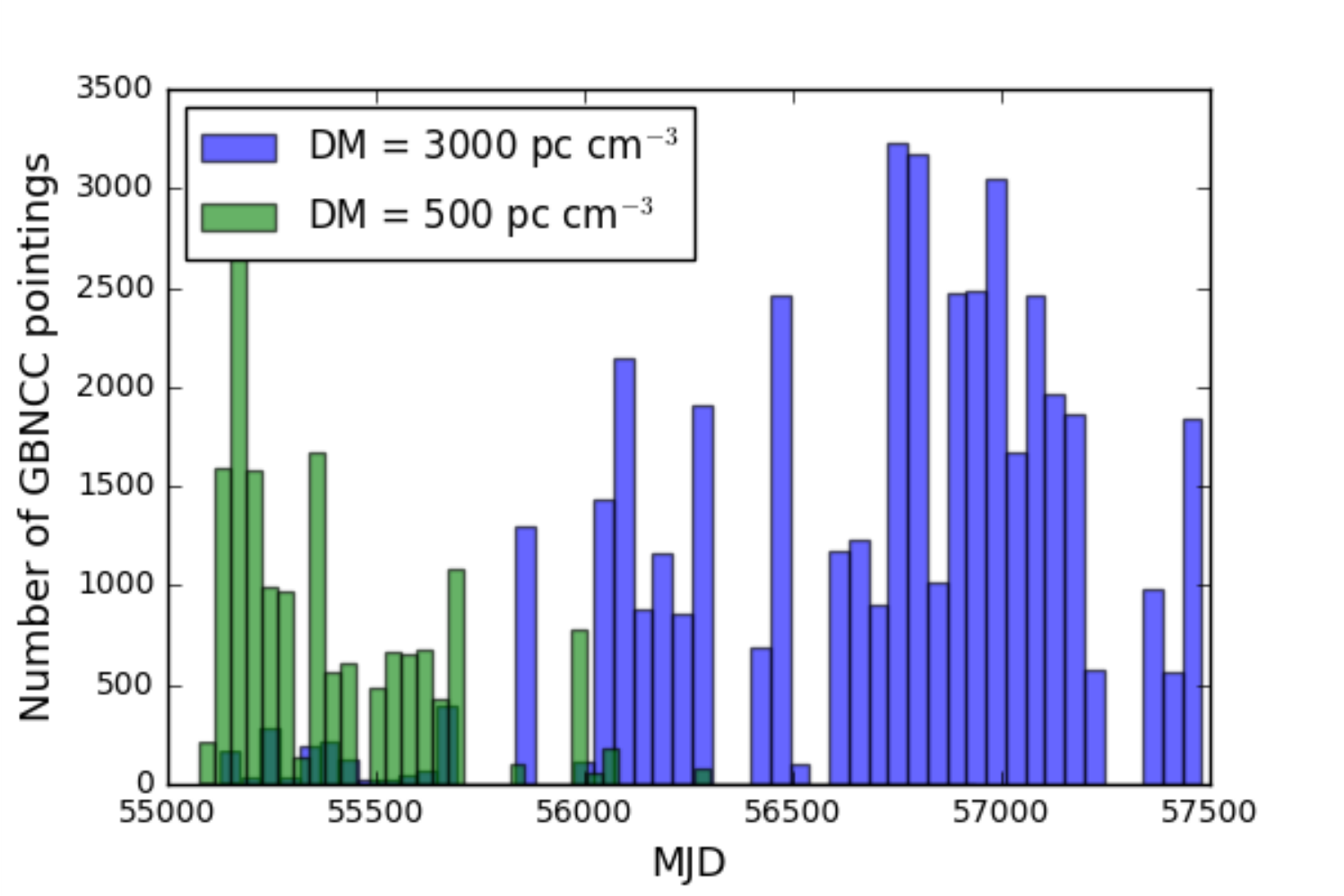}
\caption{Temporal distribution of GBNCC pointings searched for FRBs. The pointings marked in green have been searched to a DM of 500 pc cm$^{-3}$ and those in blue have been searched to a DM of 3000 pc cm$^{-3}$.}
\label{fig:time}
\end{figure}

The Green Bank Northern Celestial Cap (GBNCC) Pulsar Survey \citep{stovall2014} began in 2009 with the aim to search for pulsars and RRATs, particularly millisecond pulsars suitable for inclusion in the North American Nanohertz Observatory for Gravitational Waves (NANOGrav) pulsar timing array\footnote{\url{http://nanograv.org}}. The search is conducted using the 100-m diameter Robert C. Byrd Green Bank Telescope (GBT) at a frequency of 350 MHz. Data spanning 100 MHz of bandwidth split into 4096 frequency channels are recorded with the Green Bank Ultimate Pulsar Processing Instrument (GUPPI). Each pointing on the sky is observed for 120 s and sampled with a 81.92-$\mu$s time resolution.

The entire sky visible to the GBT ($\delta > -40 \degr$) has been divided into $\sim$125000 pointings, around 75000 of which have been observed through May 2016. In the initial days of the survey, data were searched to a maximum DM of 500 pc cm$^{-3}$. Motivated by the discovery of FRBs, the maximum DM for the search was increased to 3000 pc cm$^{-3}$. However, the initial pointings are yet to be reprocessed with this updated parameter. A total of 71\% of the pointings were searched to a DM of 3000 pc cm$^{-3}$ and 29\% of the pointings were searched to a DM of 500 pc cm$^{-3}$. The search in DM space is conducted by stepping over a range of trial DMs with $\Delta \mathrm{DM}$ being the step size between consecutive trials. The DM step sizes used by the search pipeline for the GBNCC survey are mentioned in the caption to Figure \ref{fig:smin}.

Not all pointings observed by the GBNCC survey were examined during the analysis reported on here. Pointings searched to a DM of 500 pc cm$^{-3}$ for which the maximum line-of-sight Galactic DM predicted by the NE2001 model \citep{cordes2002} was greater than 100 pc cm$^{-3}$ were not inspected. This is because our adopted definition of an FRB implies that these 7000 pointings searched over an extremely small range of extragalactic DMs as compared to the rest of the pointings. Removal of an additional 3000 pointings that were rendered unusable by the presence of radio frequency interference (RFI) limited the total observing time for the FRB search to 84 days. The time corresponding to an estimated masking fraction of 2\% for all pointings has been subtracted from the total time on sky reported here. 

Figure \ref{fig:skymap} shows the GBNCC pointings included in our FRB search overlaid on the sky map of the maximum Galactic DM predicted by the NE2001 model \citep{cordes2002}. The temporal distribution of the pointings is shown in Figure \ref{fig:time}.

\subsection{Survey Sensitivity}
 
\begin{figure*}[ht!]
\gridline{\fig{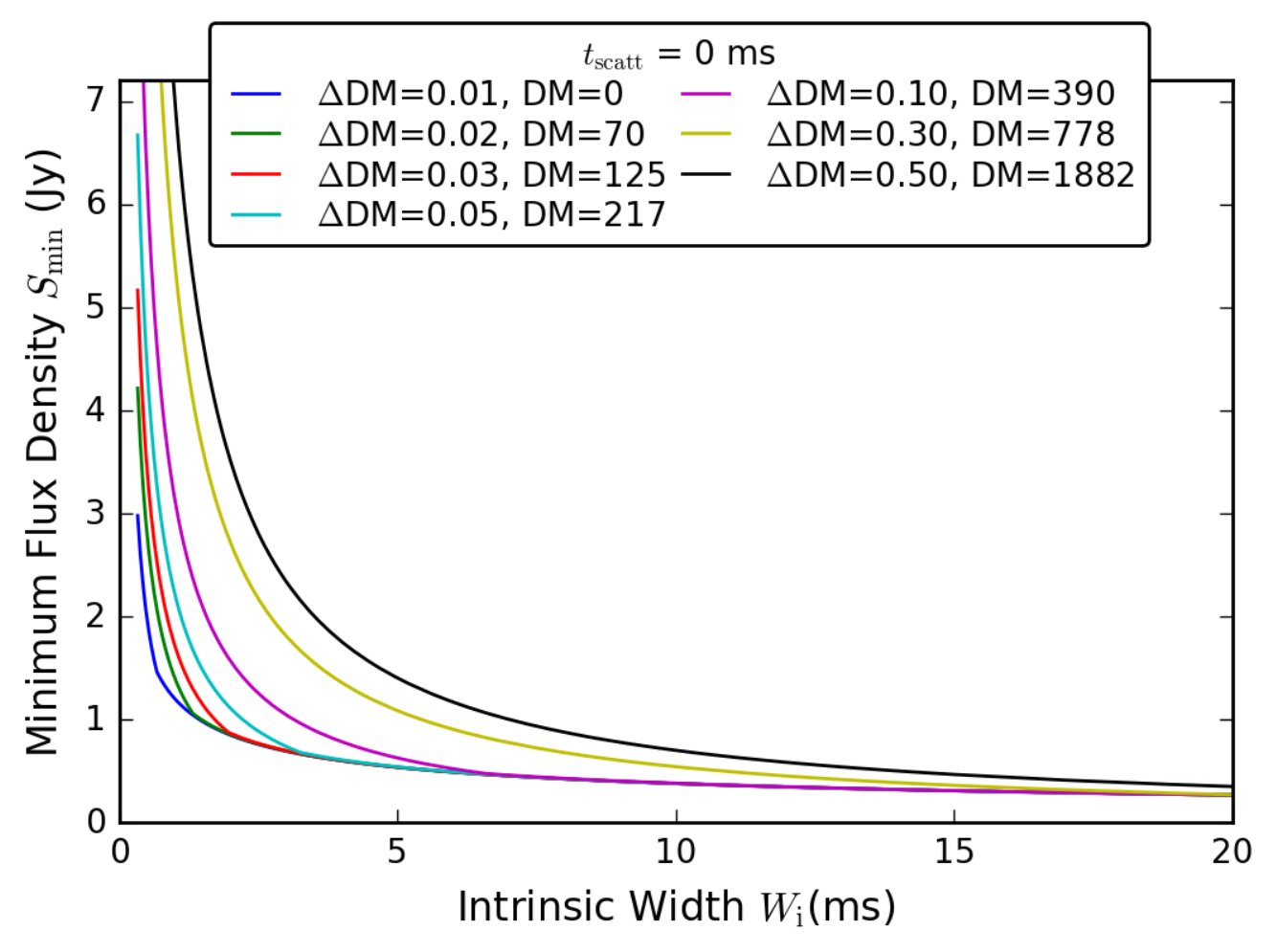}{0.5\textwidth}{(a)}
          \fig{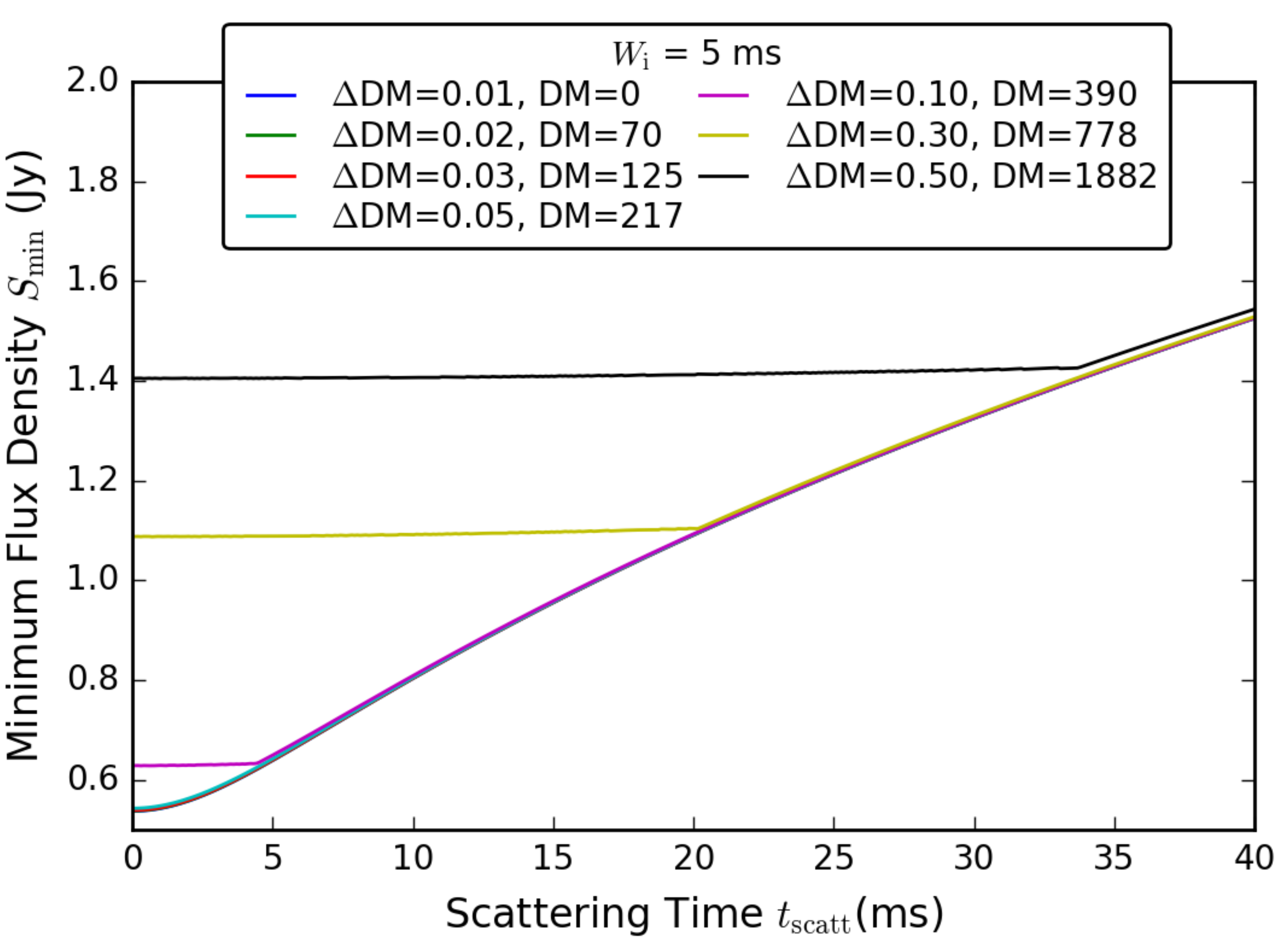}{0.5\textwidth}{(b)}
          }
\caption{Minimum detectable flux density $S_\mathrm{min}$, corresponding to DM step sizes ($\Delta \mathrm{DM}$) of 0.01, 0.02, 0.03, 0.05, 0.10, 0.30 and 0.50 pc cm$^{-3}$, plotted as a function of intrinsic pulse width $W_\mathrm{i}$ in panel (a) and scattering timescale $t_\mathrm{scatt}$ at 350 MHz in panel (b). The S/N used in the estimation of the minimum detectable flux density is a function of the pulse width $W_\mathrm{i}$, scattering timescale $t_\mathrm{scatt}$ and the DM step size, $\Delta \mathrm{DM}$. The specified DM values have units of pc cm$^{-3}$ and are the lower bounds of the trial DM ranges for the given DM step sizes.}
\label{fig:smin}
\end{figure*}

The minimum detectable flux density $S_{\mathrm{min}}$ for FRBs searched with the GBNCC survey can be calculated using the expression derived by \citet{cordes2003}:
\begin{equation}\label{eq:smin}
S_\mathrm{min} = \frac{\beta (S/N)_{\mathrm{b}} (T_{\mathrm{rec}} + T_{\mathrm{sky}})}{G W_{\mathrm{i}}} \sqrt{\frac{W_\mathrm{b}}{n_{\mathrm{p}} \Delta \nu}},
\end{equation}
where $\beta$ is a factor accounting for digitization losses, $(S/N)_{\mathrm{b}}$ is the minimum detectable signal-to-noise ratio of the broadened pulse, $T_{\mathrm{rec}}$ is the receiver temperature, $T_{\mathrm{sky}}$ is the sky temperature, $W_{\mathrm{i}}$ and $W_{\mathrm{b}}$ are the intrinsic and broadened pulse widths, respectively, $G$ is the telescope gain, $n_{\mathrm{p}}$ is the number of polarizations summed and $\Delta \nu$ is the bandwidth. Values of the above-mentioned parameters for the GBNCC survey are listed in Table \ref{tab:params}. We use $\Delta \nu$ = 75 MHz instead of the recorded bandwidth of 100 MHz to account for roll-off at the bandpass edges and for the estimated masking fraction of 5\% in the frequency domain. The average sky temperature at 350 MHz, $T_{\mathrm{sky}} = 44 \ \textrm{K}$, along the line of sight for all the pointings included in the FRB search has been estimated using the 408 MHz all-sky map \citep{remazeilles2015} and a spectral index of $-$2.6 for Galactic emission. 

The broadened pulse width $W_{\mathrm{b}}$ accounts for both instrumental and propagation effects, and is computed from the quadrature sum as follows: 
\begin{equation}\label{eq:wb}
W_\mathrm{b} = \sqrt{W_\mathrm{i}^2 + t_{\mathrm{samp}}^{2} + t_{\mathrm{chan}}^2 + t_{\mathrm{scatt}}^{2}}\ .
\end{equation}
Here $t_{\mathrm{samp}}$ is the sampling time and $t_{\mathrm{scatt}}$ is the scattering time arising from multi-path propagation of signals caused by an ionized medium. The dispersive delay within each frequency channel, $t_{\mathrm{chan}}$, is calculated (see, e.g.,\ \citealt{lorimer2005}) as follows:
\begin{equation}\label{eq:tchan}
t_{\mathrm{chan}} = 8.3 \ \mu s \bigg(\frac{\Delta \nu_\mathrm{chan}}{\textrm{MHz}} \bigg) \bigg(\frac{\nu}{\textrm{GHz}}\bigg)^{-3} \bigg(\frac{\textrm{DM}}{\textrm{pc cm}^{-3}}\bigg),
\end{equation}
where $\nu$ is the central observing frequency and $\Delta \nu_\mathrm{chan}$ is the channel bandwidth. 

For an intrinsic pulse width $W_\mathrm{i} = 5$ ms, scattering time $t_\mathrm{scatt} = 0 $ ms and a DM of 756 pc cm$^{-3}$ (mean DM of known FRBs; \citealt{petroff2015}), the minimum detectable flux density for the GBNCC survey is 0.63 Jy. We note that there is a reduction in sensitivity to high DM events since the dispersive delay for these events across a bandwidth of 100 MHz is a large fraction of the observation time per pointing. However, a significant fraction (29\%) of our pointings have been searched to a DM of 500 pc cm$^{-3}$, where this effect is not important. Also, since the highest DM observed for a known FRB is 1629 pc cm$^{-3}$ \citep{champion2016}, the sensitivity is impacted only for a small region of the parameter space.

The minimum detectable flux density is plotted as a function of intrinsic pulse width and scattering time, for different DM step sizes, in Figure \ref{fig:smin}. The minimum detectable S/N, used for the calculation of the minimum detectable flux density, is also dependent on the intrinsic pulse width, scattering timescale and DM step size. The dependence of the S/N on these variables is part of the code used to search and rank FRB candidates, RRATtrap (described in Section \ref{sec:RRAT}). The rationale for this dependence is detailed in Section \ref{sec:RRATSN}.

\section{Analysis} \label{sec:analysis}

The analysis pipeline, based on the PRESTO software package \citep{ransom2001}\footnote{\url{http://www.cv.nrao.edu/~sransom/presto}}, is run on the Guillimin High Performance Computing (HPC) cluster operated at McGill University by CLUMEQ \& Compute Canada. The first step of processing involves searching for and masking time samples and frequency channels containing RFI. The effect of dispersion is then corrected for by dedispersing the data at a large number of trial DMs up to a maximum of 500 pc cm $^{-3}$ or 3000 pc cm$^{-3}$. The dedispersed and downsampled time series for each trial DM is subsequently searched for single pulses using a matched filtering algorithm which convolves the time series with box cars having widths ranging from 81.92 $\mu$s to 100 ms. All single pulse events with S/N greater than 5 are stored for further processing. The above-mentioned analysis has been described in detail in \citet{stovall2014}. The single-pulse output is processed by a grouping and rating algorithm RRATtrap\footnote{The code is available at \url{https://github.com/ajosephy/Clustering/} and is a modified version of the code by \citet{karako2015}, which is available at \url{https://github.com/ckarako/RRATtrap}} which has aided in the discovery of 10 RRATs in GBNCC survey data \citep{karako2015}.  

\subsection{RRATtrap}\label{sec:RRAT}

\begin{figure*}[ht!]
\gridline{\fig{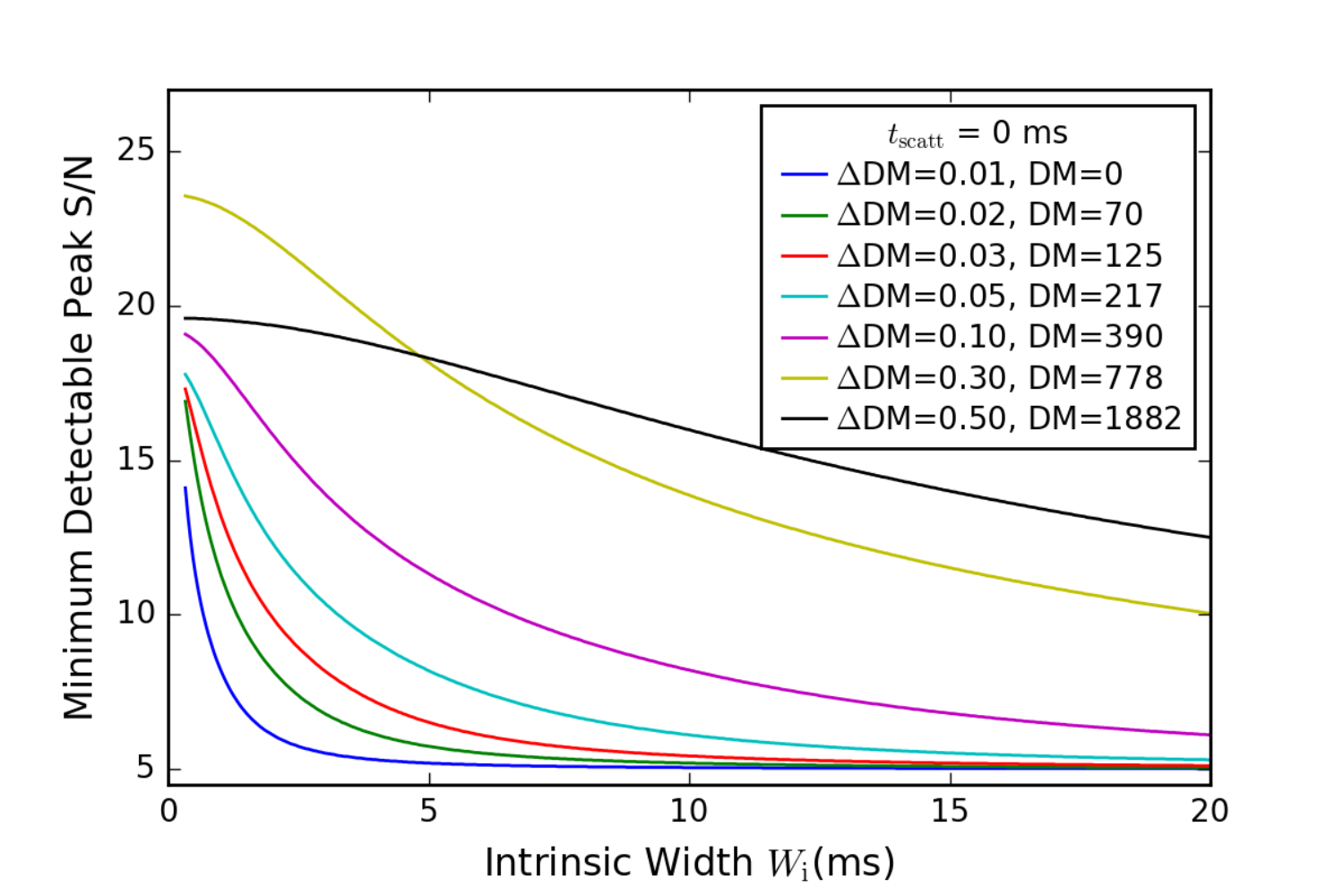}{0.5\textwidth}{(a)}
          \fig{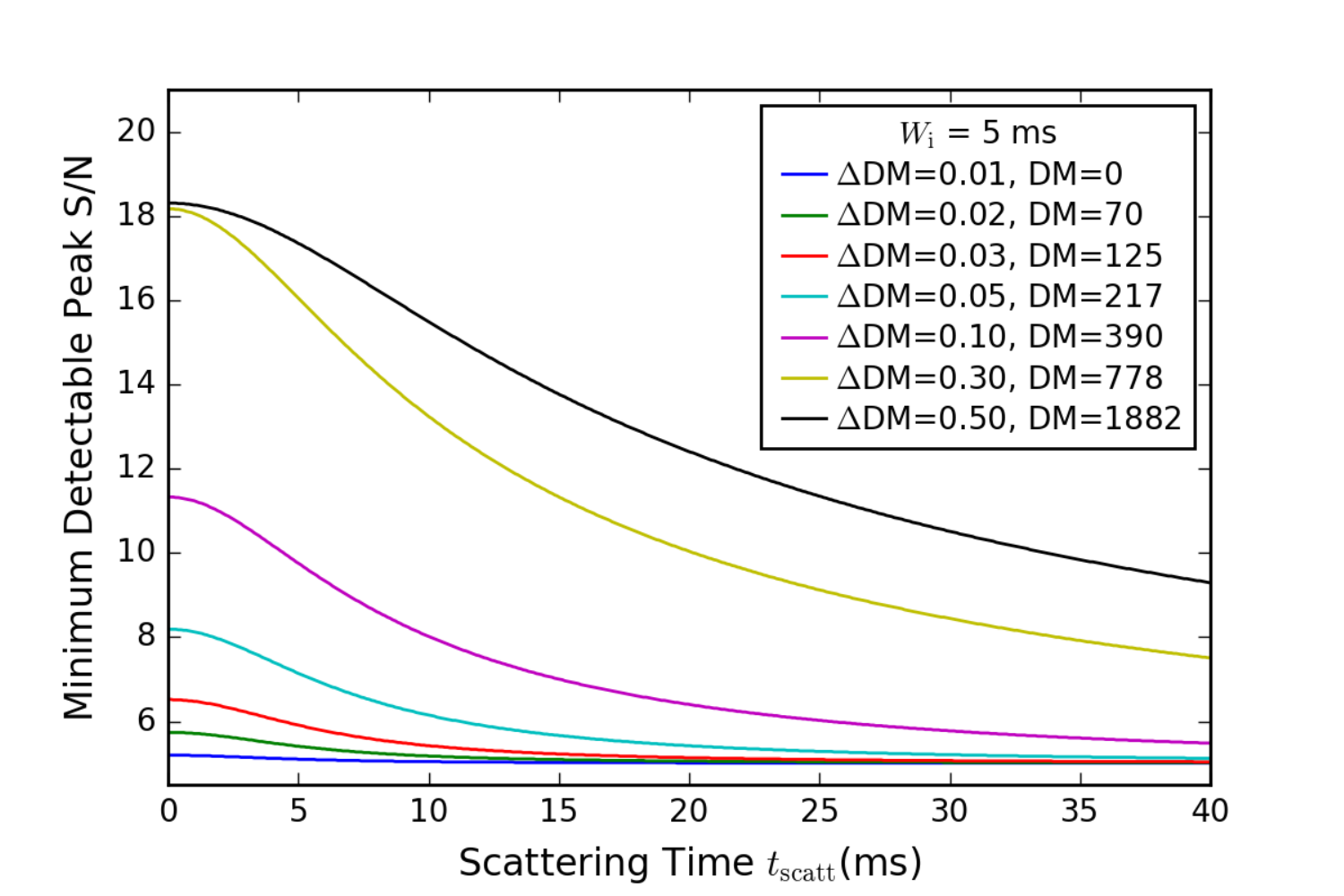}{0.5\textwidth}{(b)}
          }
\caption{Minimum detectable peak S/N with RRATtrap, corresponding to DM step sizes ($\Delta \mathrm{DM}$) of 0.01, 0.02, 0.03, 0.05, 0.10, 0.30 and 0.50 pc cm$^{-3}$, plotted as a function of intrinsic pulse width $W_\mathrm{i}$ in panel (a) and scattering timescale $t_\mathrm{scatt}$ at 350 MHz in panel (b). The specified DM values have units of pc cm$^{-3}$ and are the lower bounds of the trial DM ranges for the given DM step sizes.}
\label{fig:Svary}
\end{figure*}

The large number of DM trials ensures that each pulse (astrophysical or RFI) is detected as multiple single pulse events that are closely spaced in DM and time. RRATtrap groups all such events and ranks the groups based on how closely they match the behavior of an astrophysical pulse. It then produces colorized DM versus time plots for several DM ranges with groups of different ranks plotted in different colors. 

A group of fewer than 30 single pulse events occurring within a fixed DM and time threshold is classified as noise and not processed further. A considerable fraction of the single pulse events in our pointings fall in this category. Strong narrow-band RFI is another major source of single pulse events. The algorithm deals with these signals, that we know have a terrestrial origin, by assigning a low rank to groups with the S/N peaking at a DM $<$ 2 pc cm$^{-3}$. A low rank is also assigned to a group corresponding to a narrow-band signal, identified by it being detected with a constant S/N over a large range of DMs. A bright, broadband signal from an astrophysical source will be detected with the maximum S/N at an optimal DM and with lower S/N at closely spaced trial DMs above or below the optimal DM due to dispersive smearing. Groups exhibiting this characteristic of astrophysical pulses are ranked highly. 

\subsubsection{RRATtrap Sensitivity}\label{sec:RRATSN}
RRATtrap exhibits a significant variation in sensitivity with pulse width due to our requirement of a minimum of 30 single pulse events for a group to be ranked. Sensitivity to extremely narrow pulses is reduced since dispersive smearing prevents the detection of the pulse at 30 DM trials. The reduction in the sensitivity is maximum at high DMs where the DM step size increases to 0.5 pc cm$^{-3}$.

In order to determine whether a pulse will be ranked by RRATtrap, we first obtain the peak flux $S$ corresponding to the S/N of the pulse at the optimal DM (S/N$_\mathrm{peak}$), using Equation \ref{eq:smin}. The reduction in the peak flux $S$ of the pulse due to dedispersion at an incorrect trial DM is modelled by the following equation derived by \citet{cordes2003}:
\begin{equation}
\frac{S(\delta \mathrm{DM})}{S} = \frac{\sqrt{\pi}}{2} \zeta^{-1} \mathrm{erf} \zeta,
\end{equation}
where
\begin{equation}
\zeta = 6.91 \times 10^{-3} \delta \mathrm{DM} \frac{\Delta\nu}{W_\mathrm{i,ms} \nu^{3}_\mathrm{GHz}}.
\end{equation}
Here $\nu_\mathrm{GHz} = 0.350$ GHz is the center frequency of the GBNCC survey and $S(\delta \mathrm{DM})$ is the reduced flux measured at a trial DM differing from the optimal DM by $\delta \mathrm{DM}$. The width of the pulse dedispersed at an incorrect trial DM is given by $W(\delta \mathrm{DM}) = S W_\mathrm{i} / S(\delta \mathrm{DM})$ since dispersive smearing conserves pulse area $A = S W_\mathrm{i}$ \citep{cordes2003}. A single pulse event at a trial DM, with a DM error of $\delta \mathrm{DM}$, will therefore be detected with a S/N which can be determined by substituting the reduced flux $S(\delta \mathrm{DM})$, the intrinsic pulse width $W(\delta \mathrm{DM})$ and other parameters of the GBNCC survey in Equation \ref{eq:smin}.

For a given pulse width, we can thus obtain the minimum value of the peak S/N that will allow detection of 30 single pulse events with a S/N $> 5$. The minimum detectable peak S/N is plotted as a function of intrinsic pulse width and scattering time, for different DM step sizes, in Figure \ref{fig:Svary}. The S/N used to calculate the threshold flux density of the GBNCC survey is set to be the minimum value of the peak S/N evaluated using the above-mentioned method or 10, whichever is greater. This is done to account for the fact that only pointings having a FRB candidate with a S/N $ > 10$ were visually inspected (see Section \ref{sec:visual}).

\floattable
\begin{deluxetable}{ccccccccc}
\tablecaption{Search Parameters for Various FRB Surveys \label{tab:params}}
\tablecolumns{9}
\tablenum{1}
\tablewidth{0pt}
\tablehead{
\colhead{Survey} &
\colhead{Field of View } &
\colhead{Bandwidth} & \colhead{Center Freq.}  
& \colhead{No. of Freq.}
& \colhead{Polarizations}
& \colhead{Gain\tablenotemark{\textrm{a}}}
& \colhead{$T_\mathrm{rec}$}
& \colhead{Ref.} \\
\colhead{} & 
\colhead{(sq. deg.)} &
\colhead{(MHz)} 
& \colhead{(MHz)} 
& \colhead{Channels} 
& \colhead{Summed} 
& \colhead{(K/Jy)} 
& \colhead{(K)} 
& \colhead{}
}
\startdata
GBNCC & 0.408 & 100 & 350 & 4096 & 2 & 2 & 23 & 1  \\
PARKES\tablenotemark{b} & 0.559\tablenotemark{c} & 340 & 1352 &  1024 & 2 & 0.64 & 23 & 2 \\
UTMOST & 4.64 x 2.14 & 31.25 & 843 & 40 & 1 & 3.6 & 70 & 3 \\
PALFA & 0.022 & 322 & 1375 & 960 & 2 & \multicolumn{2}{c}{SEFD = 5} & 4 \\
CHIME & 134 & 400 & 600 & 16000 & 2 & 1.38 & 50 & 5\\   
AO327 & 0.049 & 57 & 327 & 1024 & 2 & 11 & 115 & 6 \\ 
GBT (800 MHz) & 0.055 & 200 & 800 & 4096 & 2 &  2 & 26.5 & 5 \\
LPPS (LOFAR) & 75 & 6.8 & 142 & 560 & 2 &  \multicolumn{2}{c}{SEFD = 1141} & 7\\
ARTEMIS (LOFAR) & 24 & 6 & 145 & 64 & 2 & \multicolumn{2}{c}{SEFD = 1100} & 8 \\
ALERT (APERTIF) & 8.7 & 300 & 1400  & 1024 & 2 & 0.96 & 75 & 9 \\
V-FASTR & 0.364 & 32\tablenotemark{d} & 1550 & 512\tablenotemark{d} & 2 & \multicolumn{2}{c}{SEFD = 311} & 10,11 \\
\multirow{2}{*}{MWA} & 600 & 30.72 & 155 & 24 & 2 & 1\tablenotemark{e} & 50 & 12 \\
 & 145 & 30.72 & 182 & \nodata & 2 & 1\tablenotemark{e} & 50 & 13\\
VLA & 0.283 & 256 & 1396 & 256 & 2 & \multicolumn{2}{c}{SEFD = 16} & 14\\
\enddata
\tablenotetext{\textrm{a}}{Surveys for which $T_\mathrm{rec}$ and gain ($G$) were not documented have their system equivalent flux densities (SEFD) = $ (T_\mathrm{rec} + T_\mathrm{sky}) / G$ in K/Jy reported here. $T_\mathrm{sky}$ for all other surveys has been evaluated assuming an average sky temperature of 34 K at 408 MHz and a spectral index of $-$2.6 \citep{haslam1982}.}
\tablenotetext{\textrm{b}}{The parameters are valid for the HTRU survey, for which the rate was reported by \citet{champion2016}. \citet{crawford2016} estimated the FRB rate using several Parkes surveys, the parameters for which have been reported in their paper.}
\tablenotetext{\textrm{c}}{The field of view quoted here for the 13-beam receiver of the Parkes telescope has been calculated based on the single-beam field of view of 0.043 sq. deg., reported by \citet{BSB2014}.}
\tablenotetext{\textrm{d}}{The no. of frequency channels and bandwidth reported for V-FASTR are representative values as the observing set-up can vary between observations. }
\tablenotetext{\textrm{e}}{The gain for MWA is given by A$_{\mathrm{eff}}$/2k, where k is the Boltzmann constant and A$_{\mathrm{eff}}$ is the effective area of the telescope reported by \citet{tingay2013}.} 
\tablenotetext{}{References: 1 - \citet{stovall2014}, 2 - \citet{BSB2014}, 3 - \citet{caleb2016}, 4 - \citet{scholz2016}, 5 - \citet{connor2016}, 6 - \citet{deneva2016}, 7 - \citet{coenen2014}, 8 - \citet{karastergiou2015}, 9 - \citet{vanleeuwen2014}, 10 - \citet{spolaor2016}, 11 - \citet{wayth2012}, 12 - \citet{tingay2015}, 13 - \citet{rowlinson2016}, 14 - \citet{law2015}}
\end{deluxetable}

\subsubsection{Modifications to RRATtrap}
Algorithmic changes were made to the grouping stage. Initially, this was done via ``agglomerative hierarchical clustering" (AHC) \citep{anderberg1973}, which runs in $\Oh(n^3)$ time for the simplest implementation, where $n$ is the number of single pulse events.  AHC is a bottom-up approach where all events are first initialized as individual groups and then iteratively merged based on proximity in DM and time.  Merging terminates once the minimum separation between groups, in either DM or time, is above some dimension specific threshold. The threshold in time is taken as 100 ms, corresponding to the largest boxcar used to detect pulses.  The threshold in DM is taken as 0.5 pc cm$^{-3}$ and is increased for large DMs, where the separation in trial DMs increases.

The AHC method was replaced with the ``density-based spatial clustering of applications with noise" (DBSCAN) algorithm \citep{ester1996}, which runs in $\Oh(n\log n)$ time.  
DBSCAN works by taking an arbitrary event and running a nearest neighbour query to start a group including events which are sufficiently nearby.  This group is then iteratively grown outwards by repeating the neighbourhood query for newly added members.  
Once the reachable events are exhausted, the group is complete and the process repeats for the next unvisited event. 

Since the distance thresholds used by both algorithms determine whether or not two events belong to the same group, identical thresholds yield identical output.
The purpose of the change was to reduce runtime. The performance improvement is largely due to storing the single pulse events in a $k$-d Tree \citep{bentley1975} which allows neighbourhood queries to be done in logarithmic time. 

A $k$-d Tree is a space-partitioning data structure used to organize data existing in $k$ dimensions. For our two dimensions, the tree is constructed as follows. The median event in time is taken as the root, which partitions the plane in two.  Now for each side, median events in DM are taken to further partition the plane into four regions-- these two events are the nodes in the second level of the tree. This process continues, cycling in DM and time, until all events exist as nodes on the tree. The construction of the tree takes $\Oh(n\log n)$ time.  

\subsection{Visual Inspection} \label{sec:visual}
A total of 72\% of the pointings had at least one single pulse with a S/N $> 10$. These 44000 pointings were processed with the modified version of RRATtrap to group and rank single pulse events at a DM greater than twice the maximum line-of-sight Galactic DM, DM$_\mathrm{max}$. There is a 10\% chance that an astrophysical pulse will not be ranked highly by RRATtrap \citep{karako2015}. To ensure no effect of this false negative rate on our search, we did not apply RRATtrap ranks as a criteria for visual inspection and inspected plots (corresponding to DM ranges for which DM $>$ 2DM$_\mathrm{max}$) for all 44000 pointings, regardless of the ranks of the groups they contained. However, the colors corresponding to the ranks guided the eye during the inspection of the plots. We flagged potential astrophysical candidates in these pointings and obtained their dynamic spectrum, or their intensity as a function of frequency and time. All flagged candidates had characteristics consistent with RFI and showed no evidence of a dispersive sweep. We conclude that no FRB with a S/N greater than the detection threshold of RRATtrap (see Figure \ref{fig:Svary}) was present in these pointings.

\section{Calculation of FRB Rate} \label{sec:rate}

\subsection{ Estimation of Sky Rate}
The non-detection of FRBs in our search is a significant result since it constrains the all-sky FRB rate at 350 MHz. Assuming FRBs follow Poisson statistics, the probability of detecting $N$ FRBs,
\begin{equation}\label{eq:poisson}
P(N) = \frac{ (R T \Omega)^{N} e^{-(R T \Omega)}}{{N!}},
\end{equation}
where $\Omega$ is the solid angle of the beam, $T$ is the total observing time and $R$ is the FRB rate per unit solid angle. The 95\% confidence upper limit on the rate is the upper bound for which normalization and integration of Equation \ref{eq:poisson}, with a lower bound of $R = 0$, yields a value of 0.95 for the case $N = 0$. 

We will be reporting the rate for two different beam areas, one for the field-of-view corresponding to the FWHM of the GBT beam and another for the field-of-view at the edge of which the gain is equal to 0.64 K/Jy (i.e. the Parkes 1.4-GHz on-axis gain; \citealt{BSB2014}). The former will be referred to as the FWHM case and the latter as the Parkes-equivalent case. Since all but two of the currently known FRBs have been detected using the Parkes telescope, we estimate the rate for the Parkes-equivalent case to facilitate comparison with the Parkes 1.4-GHz rate estimate (\citealt{champion2016}; \citealt{crawford2016}). Knowing that the GBT beam is well approximated by a two-dimensional symmetric Gaussian, we obtain $\Omega$ = 0.408 sq. deg. for the FWHM case ($\theta_\mathrm{0} = 36'$).\footnote{\url{https://science.nrao.edu/facilities/gbt/proposing/GBTpg.pdf}} and $\Omega = 0.672$ sq. deg. ($\theta_\mathrm{0} = 46'$) for the Parkes-equivalent case.

The total time on sky, $T$, for GBNCC pointings searched to a DM of 3000 pc cm$^{-3}$ is 61 days and, for pointings searched to a DM of 500 pc cm$^{-3}$ is 23 days. The latter pointings are sensitive only to FRBs with low extragalactic DM contributions. Thus, we are unevenly sampling the range of extragalactic DMs for the pointings we have searched implying an uneven coverage of potential FRBs. However, if we assume that all values of extragalactic DM contribution are equally likely, we can estimate an upper limit using the total observing time of 84 days that includes both pointings searched to a DM of 3000 pc cm$^{-3}$ and 500 pc cm$^{-3}$. 

For the flux limit S$_\mathrm{min}$ = 0.63 Jy corresponding to the field-of-view-averaged gain of 1.44 K/Jy for the FWHM case, we estimate a 95\% confidence upper limit on the FRB rate of 
\begin{eqnarray}
R & < & 4.98  \times 10^{3} \ \textnormal{FRBs sky}^{-1}  \ \textnormal{day}^{-1} \  \textnormal{($T$ = 61 days)} \nonumber \\
R & < & 3.62  \times 10^{3} \ \textnormal{FRBs sky}^{-1}  \ \textnormal{day}^{-1} \  \textnormal{($T$ = 84 days)} \nonumber 
\end{eqnarray}
and, for the flux limit S$_\mathrm{min}$ = 0.76 Jy corresponding to the field-of-view-averaged gain of 1.19 K/Jy for the Parkes-equivalent case, we obtain,
\begin{eqnarray}
R & < & 3.03 \times 10^{3} \ \textnormal{FRBs sky}^{-1}  \ \textnormal{day}^{-1} \  \textnormal{($T$ = 61 days)} \nonumber \\
R & < & 2.20  \times 10^{3} \ \textnormal{FRBs sky}^{-1}  \ \textnormal{day}^{-1} \  \textnormal{($T$ = 84 days)} \nonumber . 
\end{eqnarray}

The survey is ongoing with $\sim$50000 pointings left to be observed in order to cover the GBT visible sky. A non-detection in these pointings will improve the constraint on the rate to 1.98$\times 10^3$ FRBs sky$^{-1}$ day$^{-1}$ for the FWHM case. 

\subsection{Estimation of Volumetric Rate}

We can also constrain the volumetric rate of FRBs up to the redshift out to which the GBNCC survey searches. For each pointing, we are searching out to a different redshift as the DM contribution from the Galaxy varies greatly across the sky. We estimate the DM due to our Galaxy, DM$_\mathrm{MW}$, as the maximum line-of-sight DM predicted by the NE2001 model for each of our pointings. The DM contribution of the IGM can be estimated using the following equation:
\begin{equation}
\textrm{DM}_\mathrm{IGM} = \textrm{DM}_\mathrm{thresh} - \bigg(\frac{\textrm{DM}_\mathrm{host}}{z+1} + \textrm{DM}_\mathrm{MW}\bigg)
\end{equation}
Here DM$_\mathrm{thresh}$ is the maximum DM searched by the analysis pipeline, either 3000 pc cm$^{-3}$ or 500 pc cm$^{-3}$. Assuming the electron density distribution of the potential host galaxy of the FRB progenitor to be similar to that of our Galaxy, we obtain a DM contribution for the host galaxy, DM$_{\mathrm{host}}$ = 80 pc cm$^{-3}$, by averaging over the maximum DM predicted by the NE2001 model for evenly spaced lines-of-sight through our Galaxy. However, we assume DM$_{\mathrm{host}}$ as 100 pc cm$^{-3}$ for evaluating the limiting redshift of the GBNCC survey, following \citet{thornton2013}. The above assumption is to allow for a meaningful comparison with the redshifts of 0.5 to 1 inferred by \citet{thornton2013} for four FRBs discovered with the Parkes telescope. The assumption for DM$_{\mathrm{host}}$ is reduced by a factor of $(z+1)$ to facilitate comparison with the effect of DM$_\mathrm{MW}$ and DM$_\mathrm{IGM}$ \citep{ioka2003}. The reduction in the DM of the host galaxy accounts for the decrease in the observed frequency by a factor of $(z+1)$ as compared to the emission frequency of a source at a redshift $z$ and the increase in the observed dispersive delay by a factor of $(z+1)$. The limiting redshift, $z$, for each pointing can be determined using the DM-redshift relation, $\textrm{DM}_\mathrm{IGM} = 1200 \textnormal{ $z$ pc cm}^{−3}$ (\citealt{ioka2003}; \citealt{inoue2004}). We find the mean limiting redshift, $z_\mathrm{lim}$ = 1.84, for the GBNCC pointings included in our FRB search. 

We note that there are significant uncertainties in the DM-redshift relation used for the estimation of the limiting redshift. However, the relation is corroborated by the determination of the redshift of the repeating FRB121102 and the resulting estimate of the DM of its host galaxy. The DM obtained for the host galaxy, after subtracting the Galactic DM and the DM contribution estimated for the IGM using the DM-redshift relation, is equivalent to that expected from a dwarf galaxy \citep{tendulkar2017}. The observations of \citet{tendulkar2017} also imply that the assumption of DM$_{\mathrm{host}}$ = 100 pc cm$^{-3}$ could be an underestimate if FRBs preferentially exist in dwarf galaxies. The mean limiting redshift for the GBNCC pointings reduces to $z_\mathrm{lim}$ = 1.79, if we assume DM$_{\mathrm{host}}$ to be equal to the upper limit on the inferred DM of the host galaxy of FRB121102 (225 pc cm$^{-3}$). The estimate of the mean limiting redshift is thus not sensitive to the assumption for the DM contribution of the host galaxy. 

We then estimate the comoving volume surveyed by each of the pointings using the solid angle for the GBT beam at 350 MHz, $\Omega$ = 0.408 sq. deg. and assuming Planck 2015 cosmological parameters \citep{ade2015}. The total comoving volume searched by the survey is estimated by summing up the comoving volume for all the pointings and is equal to $3.8 \times 10^{11} \ \textnormal{Mpc}^3$. We note that the above estimate is an upper limit at best since the comoving volume surveyed at 350 MHz is flux-limited and cannot be correctly determined by the maximum DM searched. The intrinsic luminosity distribution of FRBs could be such that FRBs at high redshifts have flux densities less than the survey sensitivity. Additionally, pulses from high-redshift FRBs, whose intrinsic luminosity does not limit detectability, can be broadened by intra-channel and DM step smearing. Since the threshold flux density determined by Equation \ref{eq:smin} depends on the broadened pulse width which increases with increasing redshift, high-redshift FRBs with correspondingly higher DMs are harder to detect, which can also cause our survey in this volume to be flux-limited.

The upper limit on the FRB rate per unit comoving volume inferred using our upper limit on the sky rate for the FWHM case of 3.6 $\times \ 10^{3}$ FRBs sky$^{-1}$ day$^{-1}$ is 3.5 $\times \ 10^3$ Gpc$^{-3}$ yr$^{-1}$, for isotropic emission. The rate reported here is valid up to the mean limiting redshift for the GBNCC pointings, $z_\mathrm{lim} = 1.84$, and under the assumptions that the population of FRBs does not evolve with redshift and that all FRBs located at $z < z_\mathrm{lim}$ are detectable with GBNCC. The rate could be an underestimate if FRBs exhibit beamed radio emission. This is possible if FRBs are extragalactic as the extremely high implied brightness temperatures in that scenario would suggest that the emission is coherent and beamed.

\begin{figure}[t!]
\centering
\includegraphics[scale=0.55]{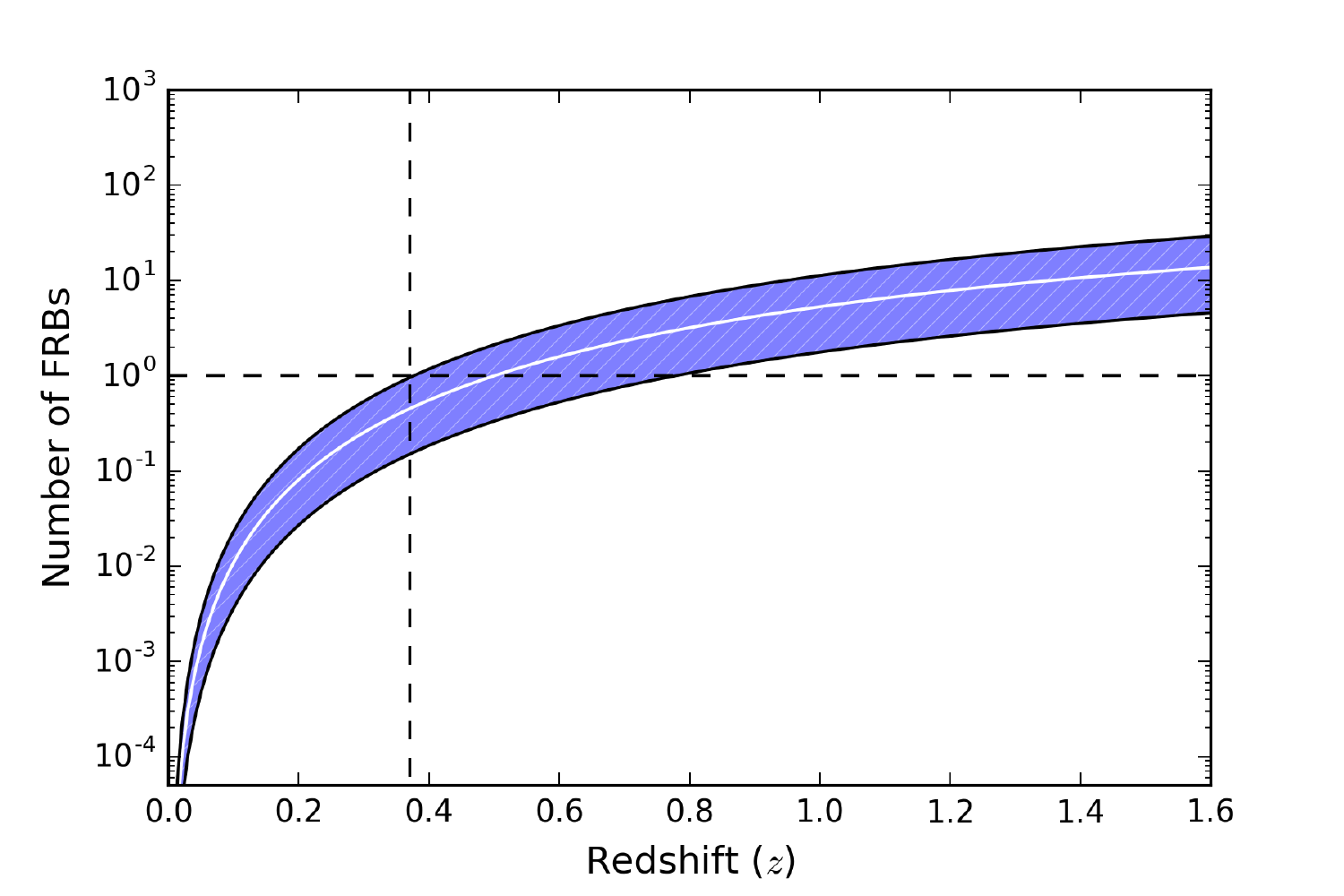}
\caption{Number of FRBs expected to be detected with the observing time and sky coverage of the GBNCC survey for different limiting redshifts. The white curve is obtained by scaling the rate reported by \citet{crawford2016} with comoving volume and the hatched region represents 99\% bounds on the rate. The GBNCC survey searched out to a limiting redshift of 0.37 as the rate for the Parkes surveys predicts detection of 1 FRB with GBNCC at $z$ = 0.37.}
\label{fig:redshift}
\end{figure}

The survey's limiting redshift and the corresponding upper limit on the volumetric rate can also be estimated following the method and assumptions detailed in \citet{rajwade2016}. The rate estimate of 3.3 $\times$ 10$^{3}$ FRBs sky$^{-1}$ day$^{-1}$ above a fluence of 3.8 Jy ms for the Parkes surveys reported by \citet{crawford2016} is assumed to be survey independent and valid at a frequency of 350 MHz and for a limiting redshift of $z_\mathrm{lim} = 0.75$ (\citealt{lorimer2013}). The limiting redshift of 0.75 is an assumption based on the redshifts of 0.5 to 1 inferred from the DMs of the FRBs discovered by \citet{thornton2013}. We translate the Parkes rate to a range of redshifts by assuming a constant comoving number density distribution of FRBs. We compute the number of FRBs detectable by GBNCC for a range of limiting redshifts using the corresponding Parkes rate. The number detectable with GBNCC for an observing time of 84 days is represented by the white curve shown in Figure \ref{fig:redshift}. The limiting redshift, $z_\mathrm{lim} = 0.37$, is the one for which the GBNCC survey is predicted to detect 1 FRB. The conclusion is justified because if the survey were sensitive to a redshift greater than $z_\mathrm{lim}$, then the Parkes rate estimate predicts a detection with the GBNCC survey, which is inconsistent with our observations. 

The above two limiting redshift estimates, obtained using different approaches, depend on several different assumptions that cannot presently be tested. The large discrepancy between the two redshift estimates can be explained if the comoving volume estimated for the first case (hereafter case A) is flux-limited such that FRBs located at $z$ $< 1.84$ are not detectable even though we are searching the DM range extending out to $z_\mathrm{lim} = 1.84$. The estimate of the limiting redshift for the second case (hereafter case B) is thus more robust since it is based on the GBNCC survey sensitivity and the underlying assumption of FRBs being standard candles which ensures that all sources in the estimated comoving volume are detectable. 

Assuming the redshift estimate of 0.37 to be correct, we conclude the upper limit on the volumetric rate to be 1.6 $\times 10^{5}$ Gpc$^{-3}$ yr$^{-1}$, with the caveat that treating the 1.4-GHz rate estimate as an all-sky rate at 350 MHz involves the implied assumption of a flat spectral index. Obtaining the rate at 350 MHz by scaling with a different assumed spectral index would change the estimate of the limiting redshift and volumetric rate. The estimate is also sensitive to the assumed intrinsic luminosity distribution and would vary if instead of the standard candle assumption, a distribution of luminosities were assumed.

\section{Spectral Index Constraints} \label{sec:spectra}
Observations of FRBs can help determine the intrinsic spectral index if the position of the FRB within the telescope beam is known. \citet{keane2016} measured $\alpha = 1.3 \pm 0.5$ for FRB150418 assuming that the FRB is located at the position of the potentially associated variable source found within the Parkes beam. The association has, however, been questioned by \citet{williams2016} and \citet{vedantham2016a}. The intrinsic spectral index can also be constrained by methods other than observation and localization. In this section, we use the non-detection with GBNCC to constrain the spectral index for different astrophysical scenarios. 

We perform Monte Carlo simulations for FRB flux distributions consistent with the rate estimate reported at 1.4 GHz for the Parkes surveys, 3.3 $\times$ 10$^3$ FRBs sky$^{-1}$ day$^{-1}$ \citep{crawford2016}. We assume a power-law flux density model for FRBs with flux density at a frequency $\nu$, $S_{\nu} \propto \nu^{\alpha}$. The cumulative flux density distribution function (the log $N$-log $S$ function) of the FRB population is also modelled as a power law with an index $\gamma$. This implies that the number of FRBs with a flux density greater than $S$, 
\begin{equation}
N ( > S) \propto S^{-\gamma}.
\end{equation}
For a non-evolving population uniformly distributed in a Euclidean universe, $\gamma = 1.5$, for any luminosity distribution. Any value other than 1.5 would argue for FRBs being a cosmological population and/or exhibiting redshift-dependent evolution. \citet{vedantham2016b} calculate $\gamma$ based on multiple-beam detections with Parkes and different detection rates for varying dish diameters, and report a constraint, 0.66 $< \gamma < $0.96. \citet{oppermann2016} derive the constraint $0.8 \leq \gamma \leq 1.7$ making use of the detections with the HTRU survey at Parkes and PALFA Survey at Arecibo. We use three different values of the slope of the log $N$-log $S$ function ($\gamma$ = 0.8, 1.2 and 1.5) for our simulations to roughly sample the range in which it is estimated to vary. 

\subsection{Absence of Scattering \& Free-Free Absorption} \label{sec:absence}

To reconcile the upper limit on the FRB rate obtained from GBNCC with the observed rate from the Parkes surveys, it may be that FRBs are rendered undetectable at low frequencies by scattering and/or the presence of a spectral turnover, either intrinsic to the emission mechanism or due to free-free absorption. In the absence of scattering and free-free absorption, the intrinsic spectral index needs to be relatively flat or even positive to account for our non-detection. 

\begin{figure}
\includegraphics[scale=0.55]{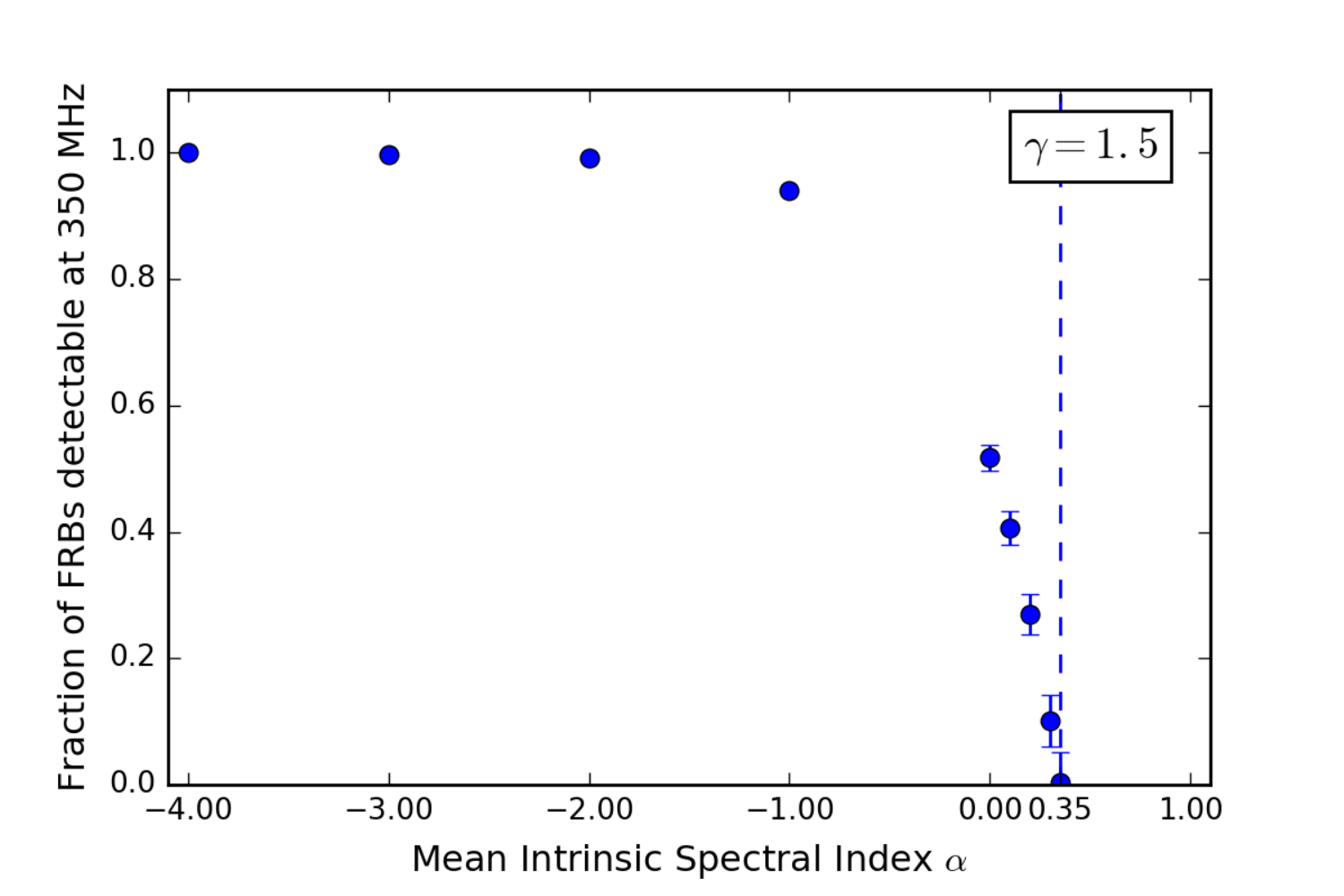}
\caption{The number of GBNCC-detectable FRBs plotted as a fraction of the computed all-sky FRB rate for GBNCC assuming a FRB population consistent with the 1.4-GHz Parkes rate estimate. Any spectral index less than the constraint, $\alpha_\mathrm{lim}$, can be rejected as it predicts FRB detections with GBNCC. The dashed line marks the constraint, $\alpha_{lim} = 0.35$, for a Euclidean flux distribution ($\gamma = 1.5$). The error bars correspond to 3$\sigma$ uncertainties and the plotted spectral indices correspond to the mean of a normal distribution with a width of 0.5.}
\label{fig:fraction}
\end{figure}

We ran 100 Monte Carlo iterations each for different cumulative flux density distributions ($\gamma$ = 0.8, 1.2 and 1.5). For each Monte Carlo iteration, we generated a flux density distribution of FRBs at 1.4 GHz consistent with the rate for Parkes surveys. We scaled the distribution to 350 MHz by sampling the spectral index of each FRB from a normal distribution ($\sigma = 0.5$) centered on the mean spectral index $\alpha$ ranging from $-4$ to $+1$. From the resulting flux distribution, we computed the all-sky rate of FRBs above a peak flux density of 0.63 Jy at 350 MHz. Figure \ref{fig:fraction} shows the number of GBNCC-detectable FRBs i.e. the difference of the computed all-sky FRB rate and the 95\% confidence GBNCC upper limit as a fraction of the computed all-sky FRB rate, for a range of spectral indices. 

The constraining spectral index is the one for which the computed all-sky rate was found to be equal to the 95\% confidence GBNCC upper limit i.e. when simulations do not predict any detections in the absence of scattering and free-free absorption. The constraints on the mean spectral index for different values of $\gamma$ are listed in Table \ref{tab:allconst}. The strongest constraint, $\alpha > 0.35$, was obtained for a Euclidean flux distribution ($\gamma$ = 1.5). The constraint depends on the assumed width of the distribution of spectral indices since the detectable FRBs in each distribution will be those with lower spectral indices. Therefore, decreasing the width will weaken the constraint on the mean spectral index. In the event of all FRBs having the same spectral index, we derive the constraint $\alpha > 0.09$, for $\gamma$ = 1.5. 

\subsection{Scattering}\label{sec:scatter}

\begin{figure}[ht!]
\centering
\includegraphics[scale=0.55]{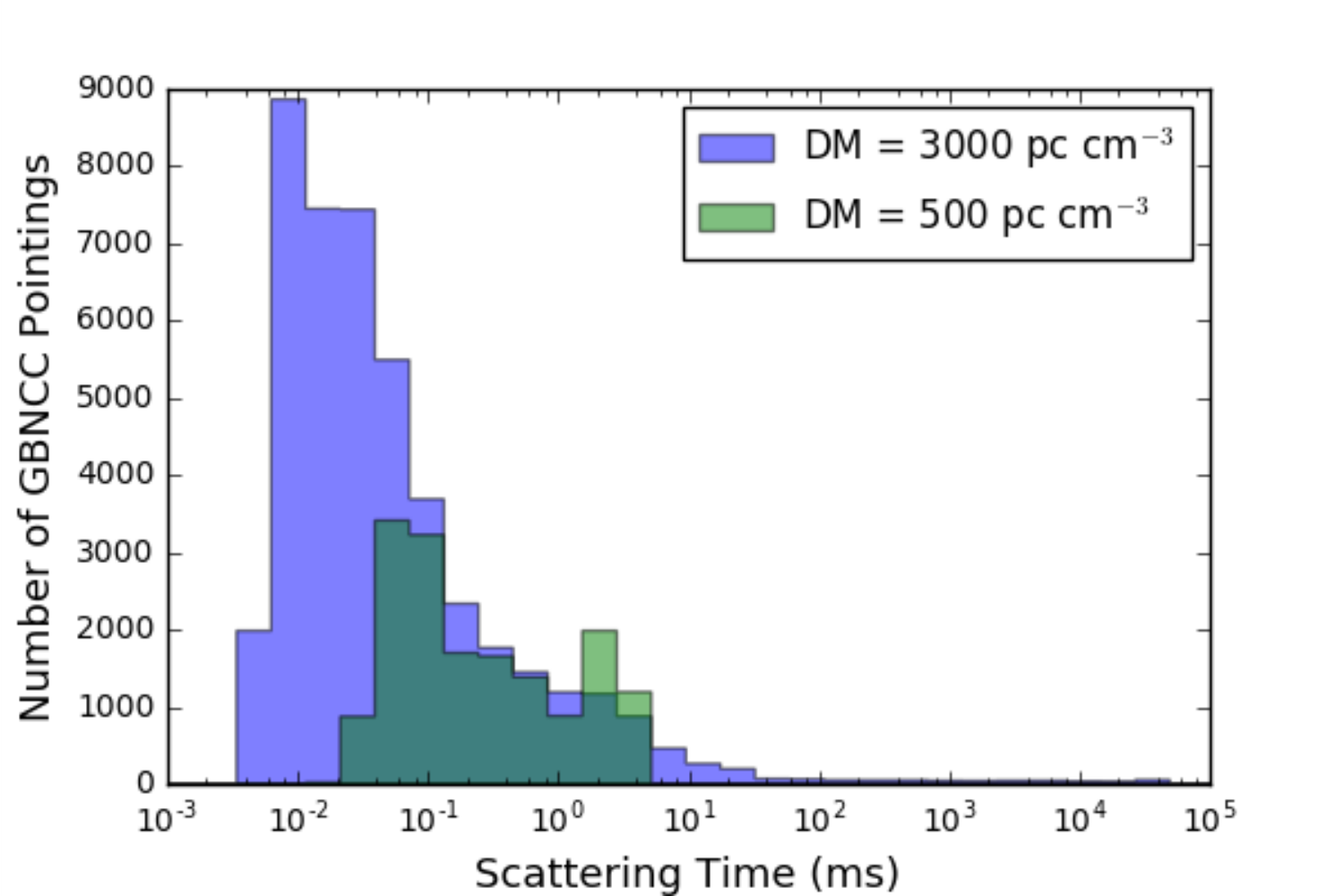}
\caption{Scattering timescales at 350 MHz predicted by the NE2001 model for all GBNCC pointings included in the FRB search. The maximum possible scattering timescales for our Galaxy along the line of sight of the GBNCC pointings are plotted here, thus assuming that the burst traverses the entire Galactic column modelled by the NE2001 model along that line of sight. The pointings in blue have been searched to a DM of 3000 pc cm$^{-3}$ and the pointings in green have been searched to a DM of 500 pc cm$^{-3}$.}
\label{fig:scatMW}
\end{figure}

Scattering may arise from three sources: our Galaxy, the intergalactic medium (IGM) and the host galaxy. Figure \ref{fig:scatMW} shows the scattering times at 350 MHz predicted by the NE2001 model along the lines of sight of all GBNCC pointings that were searched for FRBs. Since the scattering time for 98\% of these pointings is less than 10 ms (much less than our maximum searched box car width; see above), we can assume that the scattering from Galactic structures, which are modelled by the NE2001 model, is not responsible for smearing all potentially GBNCC-detectable FRBs beyond detection. However, compact regions of high electron density in our Galaxy, which are not accounted for by the NE2001 model, can potentially result in scattering timescales greater than 10 ms.

\citet{masui2015} argue against the IGM being the dominant source of scattering and support the hypothesis of strong scattering from either the dense central region of the host galaxy or a compact nebula surrounding the source. This conclusion is derived from FRB110523 showing evidence of being scattered by two plasma screens and exhibiting strong scintillation. \citet{katz2016} found no correlation between the measured pulse widths of FRBs and their extragalactic DMs suggesting that the IGM does not contribute to both scattering and extragalactic DM.  

Having established the contribution to scattering from the IGM and the Galactic structures modelled by the NE2001 model as being irrelevant for our non-detection, we ran our simulations with a three-parameter log-normal distribution of scattering times. The parameters of this distribution were chosen on the basis of the distribution of Earth-centered scattering times for our Galaxy to allow for both source models supported by \citet{masui2015}, namely a dense nebula local to the source or location in the central region of the host galaxy. The threshold parameter defines the minimum of the distribution and is set to be equal to the minimum Earth-centered scattering time for our Galaxy, 4.3$\times 10^{-3}$ ms. The scale of the distribution was set as a free parameter to allow for a range of values of the mean. The standard deviation, $\sigma = 2.74$ ms, of the underlying normal distribution was set to be the same as that of the distribution of scattering times for our Galaxy at 350 MHz predicted by the NE2001 model.

As in Section \ref{sec:absence}, we generated a flux density distribution at 1.4 GHz and scaled it to 350 MHz using spectral indices drawn from a normal distribution centered on the mean spectral index ($-4 < \alpha < 0$). We estimated the threshold flux density of the GBNCC survey to be 0.82 Jy for $t_\mathrm{scatt}$ = 10 ms, accounting for the contribution to scattering from the IGM and our Galaxy. FRBs in the flux distribution that are detectable with GBNCC (S $>$ 0.82 Jy) were assigned scattering times drawn from the above-mentioned log-normal distribution. This step was repeated with the mean of the log-normal distribution increased for each repetition until the scattering timescales of all detectable FRBs became greater than 100 ms. Since the widest box car template used by our search pipeline for detecting single pulses is 100 ms (see Section 3), FRBs with a scattering timescale greater than 100 ms will not be detected with an optimal S/N by our search pipeline. The above analysis assumes uniform sensitivity to pulses of any scattering timescale less than 100 ms.  Although there is a reduced sensitivity to highly scattered pulses because of the prevalence of RFI on longer timescales, the effect is countered by the reduction in the minimum peak S/N required to satisfy RRATtrap's cluster requirement with increase in the scattering timescale, as shown in Figure \ref{fig:Svary}.

\begin{figure}
\centering
\includegraphics[scale=0.55]{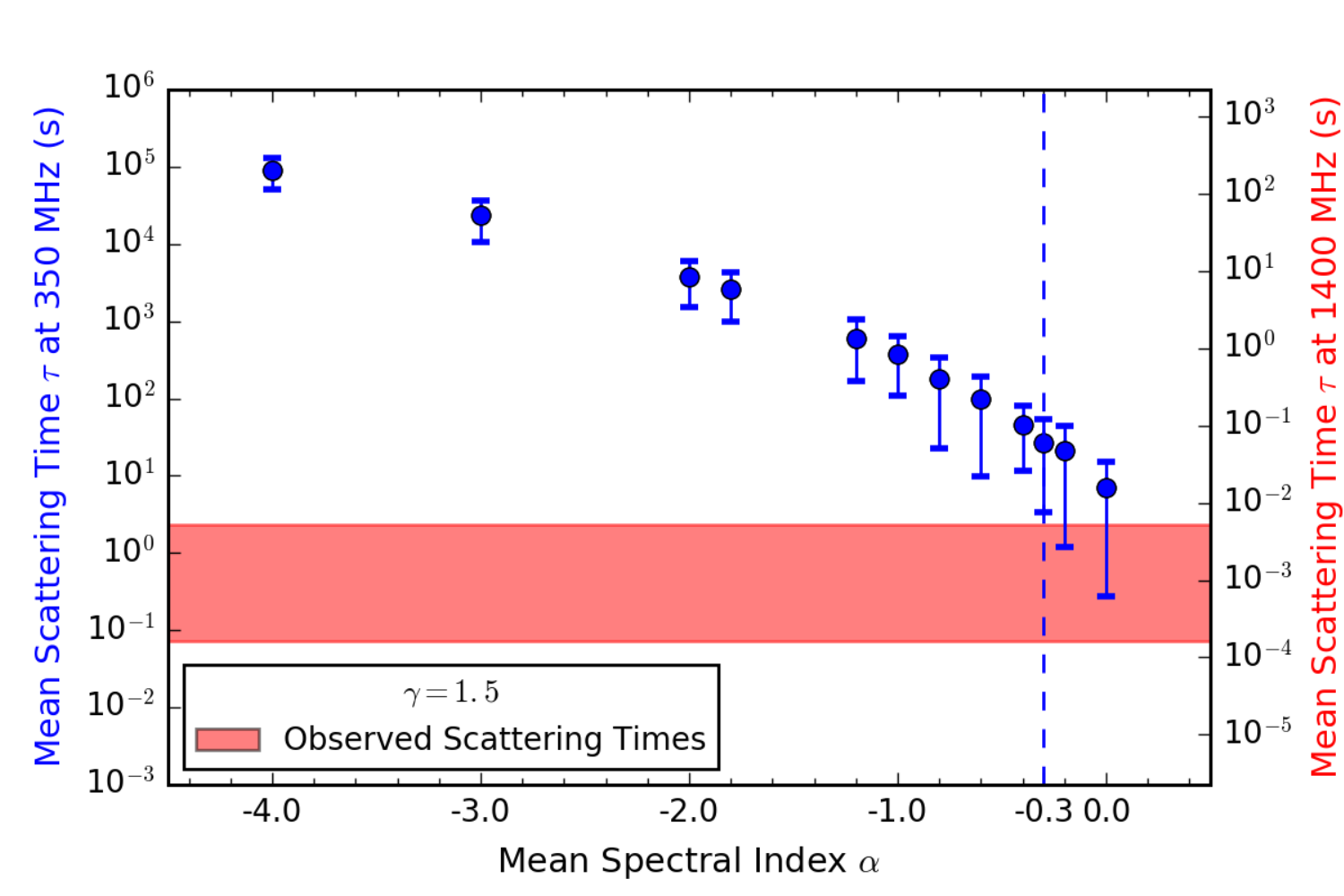}
\caption{The mean scattering time of the log-normal distribution that would render FRBs with a particular spectral index undetectable with GBNCC. The dashed line marks the spectral index constraint, $\alpha_{lim} = -0.3$, for a Euclidean flux distribution ($\gamma = 1.5$). The x-axis represents the mean of a normal distribution of spectral indices with a width of 0.5. The pink shaded band corresponds to the observed range of scattering times \citep{cordes2016} at 1.4 GHz. 2$\sigma$ error bars have been plotted. \label{fig:scatter}}
\end{figure}

Figure \ref{fig:scatter} shows the mean scattering time of the log-normal distribution that can render FRBs in the flux density distribution expected to be seen by GBNCC (with S $>$ 0.82 Jy) in an observing time of 84 days undetectable, for a range of spectral indices. A more negative spectral index would predict a higher number of detections with GBNCC requiring a higher mean scattering time at 350 MHz to render all the FRBs undetectable. We find our constraint on the spectral index, $\alpha_\mathrm{lim}$, to be the one for which the mean scattering time at 350 MHz scales to the maximum observed scattering timescale for known FRBs at 1.4 GHz assuming a Kolmogorov scaling. We derive the constraint, $\alpha_\mathrm{lim} > -0.3$, for a Euclidean flux distribution. This constraint is valid only in the absence of free-free absorption. The constraints for other values of $\gamma$ are listed in Table \ref{tab:allconst}. 

\begin{deluxetable}{cCCC}
\tablecaption{Spectral Index Constraints\label{tab:allconst}}
\tablecolumns{4}
\tablenum{2}
\tablewidth{0pt}
\tablehead{
\colhead{$\gamma$} &
\colhead{No Scattering/FF\tablenotemark{\textrm{a}}} &
\multicolumn2c{Scattering\tablenotemark{\textrm{b}}} \\
\colhead{} & \colhead{} &
\colhead{\citeauthor{crawford2016}} & \colhead{\citeauthor{champion2016}}
}
\startdata
0.8 & > 0.19 & > -0.9 & > -1.5\\
1.2 & > 0.28 & > -0.6 & > -1.2\\
1.5 & > 0.35 & > -0.3 & > -0.9 
\enddata
\tablenotetext{\textrm{a}}{FF refers to free-free absorption.}
\tablenotetext{\textrm{b}}{The two columns correspond to different 1.4 GHz rate \\ estimates assumed for the initial flux distribution.}
\end{deluxetable}

\vspace*{-10mm}

Different surveys conducted at 1.4 GHz with the Parkes telescope have different reported rate estimates and flux density thresholds. To gauge the sensitivity of our results to the assumed 1.4-GHz rate estimate, we repeat this analysis with a flux distribution at 1.4 GHz that is consistent with the rate reported by \citet{champion2016} of 7 $\times 10^3$ FRBs sky$^{-1}$ day$^{-1}$ above a flux density of 0.17 Jy for $W_\mathrm{i}$ = 5 ms. The resulting constraints are weaker and are listed in Table \ref{tab:allconst}. The constraints on spectral index are also sensitive to the width of the log-normal distribution. Decreasing the width of the distribution would allow even modest scattering times to explain our non-detection and thus weakening the constraints on spectral index. 

Another effect which can potentially weaken our constraints is the 1.4-GHz observation of a reduced FRB detection rate at low and intermediate Galactic latitudes as compared to high Galactic latitudes by \citet{petroff2014}. A recent analysis by \citet{vanderwiel2016} demonstrates that the reduction in the FRB rate is significant (p = $5 \times 10^{-5}$) for low Galactic latitudes ($|\mathrm{b}| < 5\degr$) while the difference between the mid-latitude ($5\degr < |\mathrm{b}| < 15\degr$) and the high-latitude FRB rate is only marginally significant (p = 0.03). Since only 5\% of the GBNCC survey pointings were observed at Galactic latitudes $|\mathrm{b}|<5\degr$, and 10\% of the pointings were observed at intermediate Galactic latitudes, incorporating the latitude dependence of the FRB rate would not have a significant effect on the resulting constraints. Additionally, if the latitude dependence of the FRB event rate is due to diffractive scintillation as suggested by \citet{macquart2015} then the frequency dependence of this effect can also weaken the spectral index constraints. However, we do not account for this effect here since \citet{scholz2016} demonstrate that the analysis by \citet{macquart2015} is incorrect as its prediction for a high FRB rate with the PALFA survey is not matched by observations. 

\subsection{Constant Comoving Number Density Distribution}

We attempt to constrain the spectral index for the specific case of a constant comoving number density distribution in this section. The approach follows from the analysis in \citet{rajwade2016} and references therein and enables us to derive constraints for a variety of astrophysical models. It is based on the assumption that FRBs are standard candles, thus making it different from the approach described in Section \ref{sec:scatter}.

The bolometric luminosity, $L$ for each model and spectral index $\alpha$ is evaluated using the following equation assuming a $S_\mathrm{peak}$ = 1 Jy detection of an FRB located at $z_\mathrm{lim}$ = 0.75 \citep{lorimer2013} with the Parkes surveys:
\begin{equation} \label{eq:speak}
S_\mathrm{peak} = \frac{L \int_{\nu'_{1}}^{\nu'_{2}}{E_{\nu'}} d \nu' }{(1+z)^{2} 4 \pi D(z)^2 (\nu_2 - \nu_1) \int_{\nu'_\mathrm{low}}^{\nu'_\mathrm{high}}{E'_{\nu'}} d \nu'}.
\end{equation}
Here $D(z)$ is the comoving distance calculated using Planck 2015 cosmological parameters \citep{ade2015} and $\nu' = (1+z) \nu $ is the frequency in the source frame. The limiting frequencies for emission, $\nu'_\mathrm{high}$ and  $\nu'_\mathrm{low}$, are assumed to be 10 GHz and 10 MHz, respectively \citep{lorimer2013}. The frequencies $\nu_1$ and  $\nu_2$ are the lowest and highest observing frequencies of the survey in consideration. 

The bolometric luminosity is different for each model because of the difference in the expression for the energy released per unit frequency interval, $E_{\nu'}$. In the absence of scattering and free-free absorption, positive spectral indices will be the sole reason for reduction of flux at low frequencies and we can set $E_{\nu'} \propto \nu'^{\alpha}$. Mirroring the terminology used by \citet{rajwade2016}, we will be referring to it as model A hereafter. For the model where scattering becomes important (model B), $E_{\nu'}$ gets reduced by a factor of 
$\sqrt{1 + {(t_\mathrm{scatt}/W_\mathrm{i})}^2}$. Here $t_\mathrm{scatt}$ is the scattering timescale at a frequency $\nu$ obtained by scaling the mean observed timescale of 6.7 ms at 1 GHz under the assumption of a Kolmogorov scattering spectrum. The observed scattering time of 6.7 ms was determined by taking the average of the scattering timescales of known FRBs \citep{cordes2016}. For FRBs with no measured scattering timescales, we used half of the published upper limits when computing the average.

Another astrophysical phenomenon that can render FRBs undetectable at low frequencies is free-free absorption in the dense environment surrounding the FRB progenitor. For the case of free-free absorption, 
\begin{equation}\label{eq:ffabsorb}
E_{\nu'} \propto \ \bigg(\frac{\nu'}{1 \ \textrm{GHz}}\bigg)^{\alpha} \ \exp \Bigg[- \tau \ \bigg(\frac{\nu'}{1 \ \textrm{GHz}}\bigg)^{-2.1}\Bigg].
\end{equation}
The optical depth, $\tau$ at 1 GHz is computed using $\tau$ = 0.082 T$_\mathrm{e}^{-1.35}$ EM \citep{mezger1967}, where T$_\mathrm{e}$ is the electron temperature and EM is the emission measure. We considered two models of free-free absorption; cold molecular clouds with ionization fronts (model C) and hot, ionized magnetar ejecta/ circum-burst medium (model D). The parameters, T$_\mathrm{e}$ and EM for these models have been adopted from \citet{rajwade2016} and are listed in Table \ref{tab:comoving}. Model E and F mimic model C and D, respectively, but also account for scattering. For this, the expression for $E_{\nu'}$ in Equation \ref{eq:ffabsorb} is reduced by a factor of $\sqrt{1 + {(t_\mathrm{scatt}/W_\mathrm{i})}^2}$, as was done for model B. 

\begin{figure}[ht!]
\centering
\includegraphics[scale=0.55]{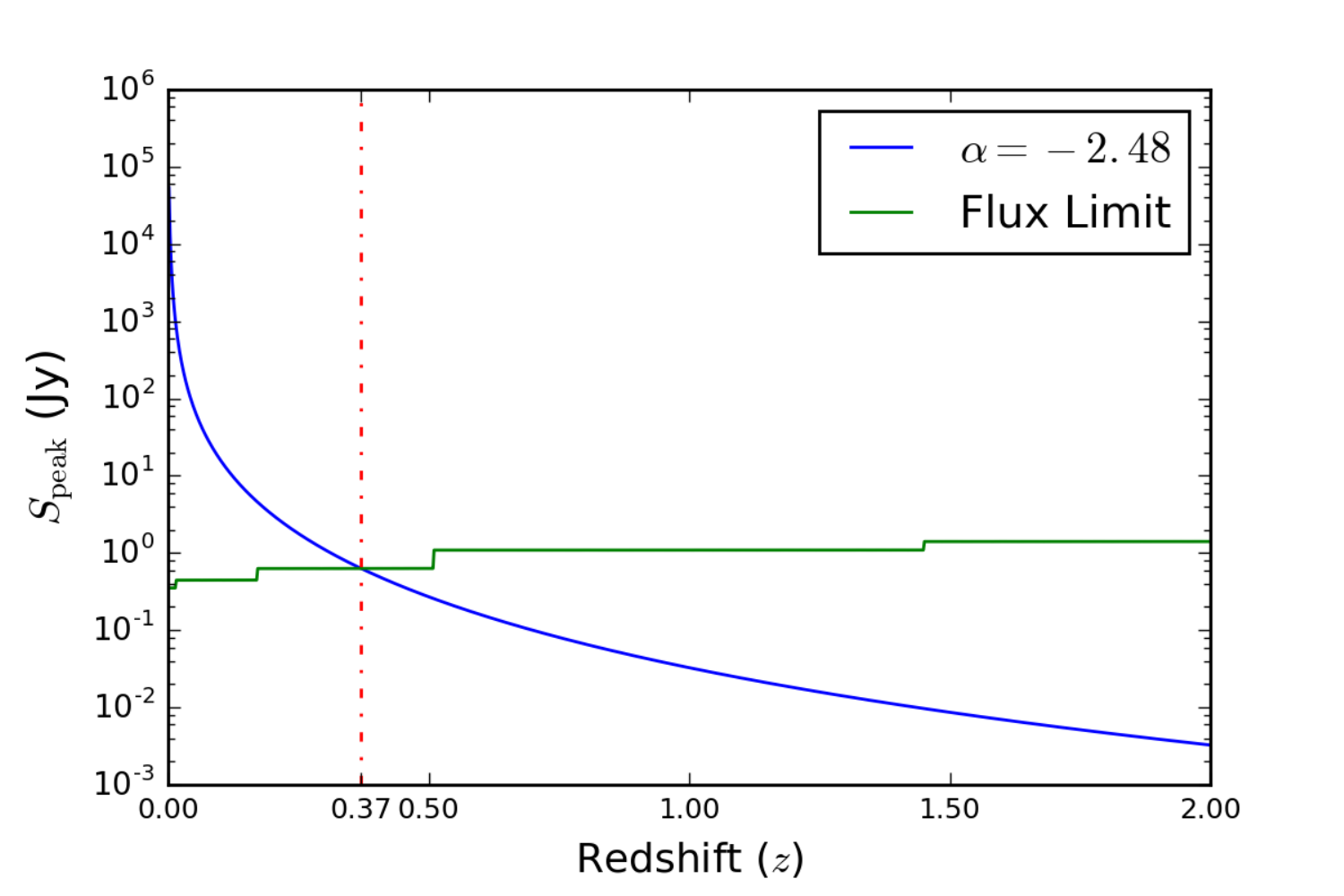}
\caption{Peak flux density as a function of redshift for the constraining spectral index, $\alpha_\mathrm{lim}$, for model B (scattering). The green line corresponds to the minimum detectable flux density for the GBNCC survey and the red line marks the limiting redshift for the GBNCC survey, $z_\mathrm{lim} = 0.37$, for case B. The peak flux density for the constraining spectral index evaluated using Equation \ref{eq:speak} and represented by the blue curve is shown to be equal to the survey sensitivity at the limiting redshift.}
\label{fig:spectral}
\end{figure}

Using the expressions for $E_{\nu'}$ derived above, we calculate the peak flux density detectable with the GBNCC survey for each model and spectral index $\alpha$ by substituting the bolometric luminosity for that model and spectral index, and parameters of the GBNCC survey in Equation \ref{eq:speak}. This calculation is performed for a range of redshifts. The peak flux density at GBNCC survey's limiting redshift for case B, $z_\mathrm{lim} = 0.37$, should be equal to the survey's sensitivity for the constraining spectral index $\alpha_\mathrm{lim}$. Any spectral index $\alpha < \alpha_\mathrm{lim}$ can be rejected as the peak flux density at $z_\mathrm{lim}$ for $\alpha < \alpha_\mathrm{lim}$ would be greater than the survey sensitivity implying FRB detections with the GBNCC survey. The procedure is shown graphically in Figure \ref{fig:spectral} and the resulting constraints are listed in Table \ref{tab:comoving}. 

The GBNCC survey sensitivity exhibits a non-linear dependence on redshift. If the IGM is assumed to be the dominant contributor to the DM, then the DM-redshift relation (\citealt{ioka2003}; \citealt{inoue2004}) implies that we are searching for FRBs with higher DMs as we search out to higher redshifts. The increase in DM increases the dispersive smearing within each frequency channel (evaluated using Equation \ref{eq:tchan}), thereby broadening the pulse and increasing the minimum detectable flux density of the survey.  The survey sensitivity for the broadened pulse width is determined using Equation \ref{eq:smin} and is plotted in Figure \ref{fig:spectral}. 

The constraints listed in Table \ref{tab:comoving} are based on the limiting redshift for the GBNCC survey for case B, the calculation of which is based on the assumption that the Parkes surveys searched to a redshift, $z_\mathrm{lim}$ = 0.75. If the repeating FRB121102 is not representative of the FRB population i.e. not all FRBs are cosmological, then this assumption might not hold true. Other caveats associated with these constraints have been detailed in \citet{rajwade2016}.  

For the case of scattering, the constraint for a uniform distribution in comoving volume of FRBs, $\alpha_\mathrm{lim} = -2.48$ is very weak in comparison with $\alpha_\mathrm{lim} = -0.3 $ evaluated for a Euclidean flux distribution with the approach described in Section \ref{sec:scatter}. The constraint for a Euclidean flux distribution is derived by assuming a distribution of scattering times as compared to a single scattering time for all FRBs assumed for evaluating the weaker constraint. The marked difference in the resulting constraints points to the sensitivity of our results to the initial assumptions about the scattering timescale. 

The constraints based on the GBNCC non-detection are not markedly different from the constraints evaluated by \citet{rajwade2016} based on non-detection with surveys such as AO327 \citep{deneva2016}, LOFAR \citep{karastergiou2015} and UTMOST \citep{caleb2016}. The constraint we derive under the assumption of absence of scattering and free-free absorption, $\alpha_\mathrm{lim} = 1.18$, is stronger than the most constraining spectral index obtained from the above-mentioned surveys ($\alpha_\mathrm{lim} = 0.7$; AO327). The best constraint derived by \citet{rajwade2016} for the model where scattering becomes relevant, $\alpha_\mathrm{lim} = -2.10$, is based on non-detection with UTMOST and is stronger than the constraint obtained using the GBNCC non-detection, $\alpha_\mathrm{lim} = -2.48$.

\begin{deluxetable}{ccCRC}
\tablecaption{\centering{Spectral Index Constraints for Constant Comoving Number Density Distribution}\label{tab:comoving}}
\tablenum{3}
\tablewidth{0pt}
\tablehead{
\colhead{Model} &
\colhead{T$_\mathrm{e}$} &
\colhead{EM} & \colhead{$\alpha_\mathrm{lim}$} \\
\colhead{} & \colhead{(K)} &
\colhead{(cm$^{-6}$ pc)} & \colhead{}
}
\startdata
A & \nodata & \nodata & 1.18\\
B & \nodata & \nodata & -2.48\\
C  & 200 & 1000 & 1.00\\
D  & 8000 & 1.5 \times 10^{6} & -0.64\\
E & 200 & 1000 & -2.67\\
F & 8000 & 1.5 \times 10^{6} & -4.39\\
\enddata
\end{deluxetable}

\section{Implications for Other Surveys}\label{sec:implication}

We can predict the FRB detection rates for current and upcoming surveys using the constraints on spectral index derived from GBNCC. We derive the following equation for the calculation of the FRB rate, $R$ above a flux density $S_\mathrm{0}$ at a frequency $\nu_\mathrm{0}$, for a spectral index $\alpha$ and slope of the log $N$-log $S$ function, $\gamma$:
\begin{eqnarray}\label{eq:rate}
R( > S_\mathrm{0} ) & = & R_\mathrm{ref} \bigg(\frac {S_\mathrm{0}} {S_\mathrm{ref} \big( \frac{\nu_\mathrm{0}} { \nu_\mathrm{ref}}\big)^{\alpha} }\bigg)^{-\gamma} \nonumber \\
 & = & R_\mathrm{ref}  \bigg(\frac{S_\mathrm{0}}{S_\mathrm{ref}}\bigg)^{-\gamma} \bigg(\frac{\nu_\mathrm{0}}{\nu_\mathrm{ref}}\bigg)^{\alpha\gamma}.
\end{eqnarray}
Here $R_\mathrm{ref}$ is the reference rate estimate above a flux density $S_\mathrm{ref}$ at a frequency $\nu_\mathrm{ref}$. The above equation uses a scaling factor of $\nu^{\alpha\gamma}$ to calculate the FRB rate instead of $\nu^\alpha$ used by \citet{BSB2014}. The correction to the scaling factor can be justified in the following manner. If $R_\mathrm{ref}$ is the number of bursts detectable per sky per day above a flux density $S_\mathrm{ref}$ at a frequency $\nu_\mathrm{ref}$, then $R_\mathrm{ref}$ is also the number of bursts detectable above a flux density $ S_\mathrm{ref} \big(\frac{\nu_\mathrm{0}}{\nu_\mathrm{ref}}\big)^\alpha$ at a frequency $\nu_\mathrm{0}$. The ratio of the number of bursts $R$ detectable above a flux density $S_\mathrm{0}$ and the number of bursts $R_\mathrm{ref}$ detectable above a flux density $S_\mathrm{ref} \big(\frac{\nu_\mathrm{0}}{\nu_\mathrm{ref}}\big)^\alpha$ can then be given by Equation \ref{eq:rate}.

However, Equation \ref{eq:rate} makes incorrect assumptions about the FRB population in that it does not allow a distribution of spectral indices and scattering timescales. This warrants the need to run Monte Carlo simulations to ensure that the predicted rate accounts for the occasional bright FRBs with scattering times lower than the mean of the population. For instance, all FRBs with a scattering time greater than 1 ms at 1 GHz will not be detectable with an optimal S/N with the GBNCC survey. This is because the widest box car template of 100 ms used by our search pipeline corresponds to a timescale of 1 ms at 1 GHz (under the assumption of a Kolmogorov medium). However, scattering timescales of known FRBs at 1 GHz range from 0.7 to 23 ms with several of these measurements being upper limits \citep{cordes2016}, suggesting that the survey could still be sensitive to a significant fraction of the FRB population.

We generated flux distribution of FRBs at 1.4 GHz consistent with the Parkes rate estimate for $\gamma$ = 0.8, 1.2 and 1.5. Spectral indices drawn from a normal distribution ($\sigma = 0.5$) centered on the mean spectral index were used to scale the flux distribution to the frequency of the survey in consideration. For each FRB in the distribution, the scattering time $t_\mathrm{350}$ was sampled from a log-normal distribution at 350 MHz with the width same as our Galaxy's distribution. The mean of the log-normal distribution was set to be the scattering timescale at 350 MHz obtained by scaling the mean observed timescale of 6.7 ms at 1 GHz assuming a Kolmogorov spectrum. The flux of all FRBs in the distribution was reduced by $\sqrt{1+ ({t_\mathrm{scatt}/W_\mathrm{i}})^2}$ where $t_\mathrm{scatt}$ was the scattering time for each FRB at the survey frequency, obtained by scaling $t_\mathrm{350}$, again under the assumption of a Kolmogorov spectrum. The number of FRBs in this distribution with a flux greater than $S_\mathrm{0}$ was used to compute the number of bursts per hour detectable by the survey. The minimum detectable flux density for each survey, $S_\mathrm{0}$, was evaluated using Equation \ref{eq:smin} with (S/N)$_\mathrm{b}$ = 10, $t_\mathrm{scatt}$ = 0 and the parameters for the surveys considered listed in Table \ref{tab:params}. 

Our simulations predict the rate for mean spectral indices ranging from $\alpha_\mathrm{lim}$ to an arbitrary upper limit, $\alpha = +2$. If $\alpha > +2$, then the rate predictions for all surveys at frequencies $<$ 1.4 GHz would decrease, although there is no observational evidence arguing for $\alpha > +2$. The lower limits on the mean spectral index, $\alpha_\mathrm{lim}$ are the constraints we obtain with GBNCC in the event of scattering which are listed in Table \ref{tab:allconst}. 

\begin{figure*} 
\vspace*{-12mm}
\gridline{\fig{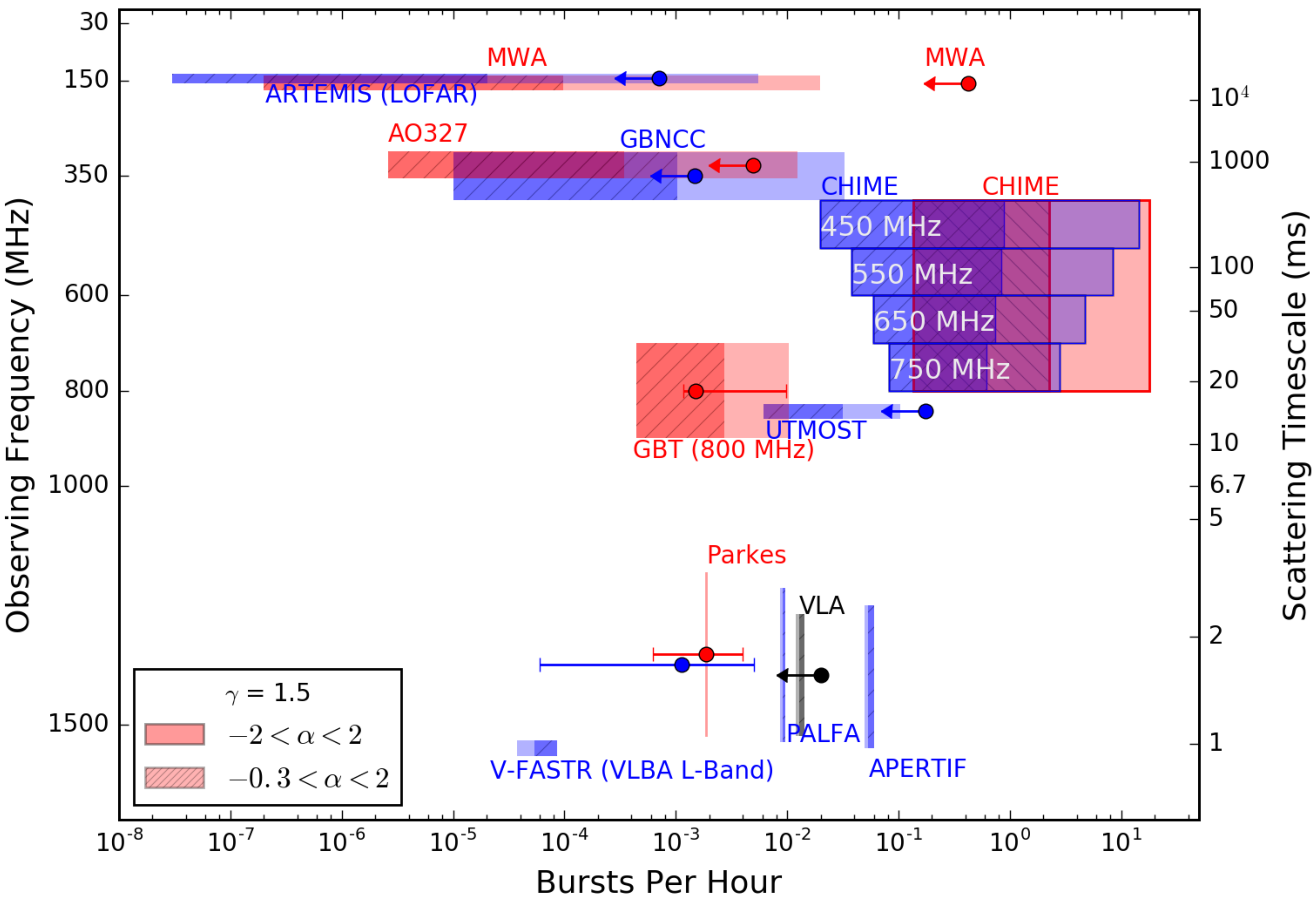}{0.55\textwidth}{(a)}}
\vspace*{-4mm}
\gridline{\fig{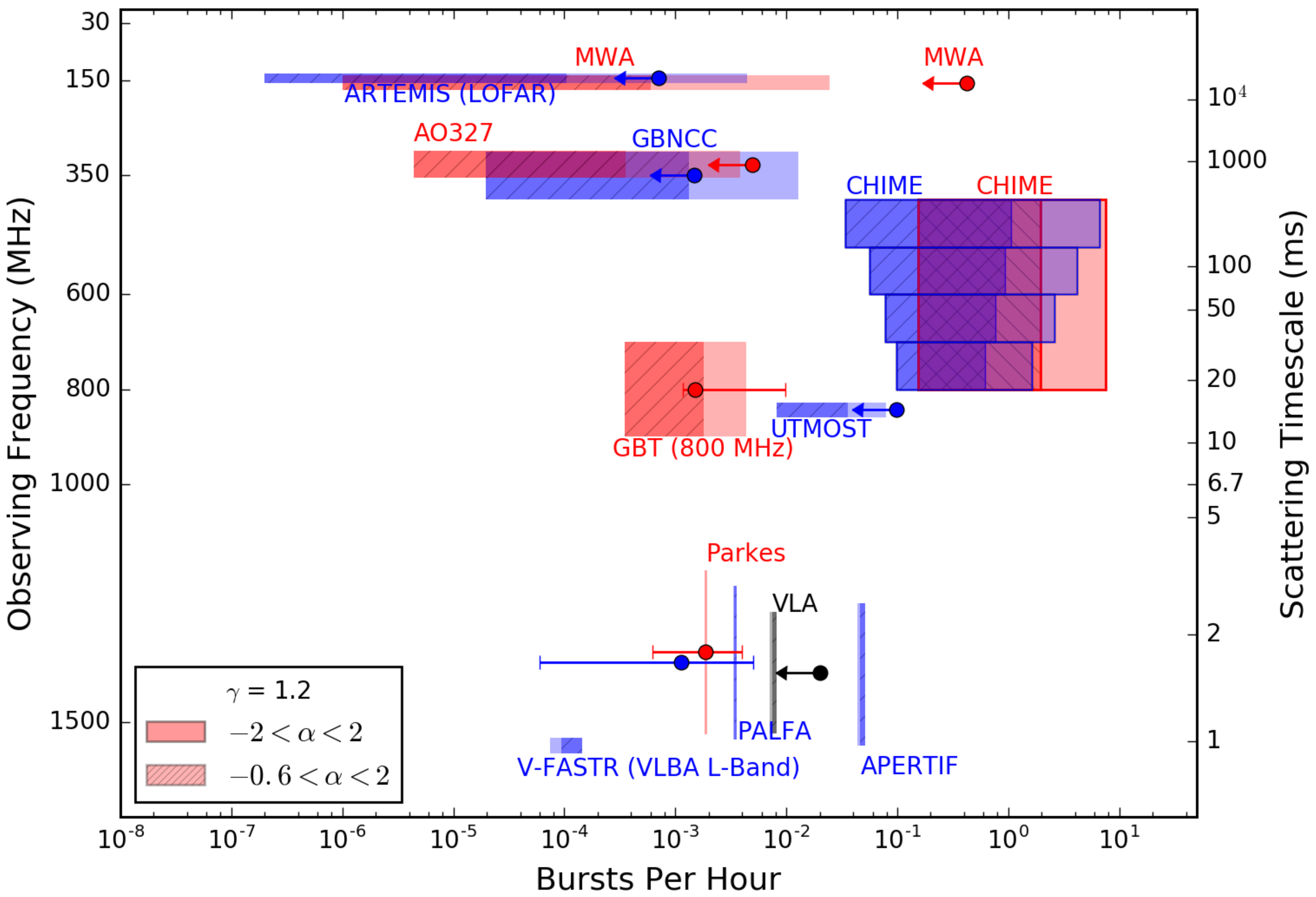}{0.55\textwidth}{(b)}}
\vspace*{-4mm}
\gridline{\fig{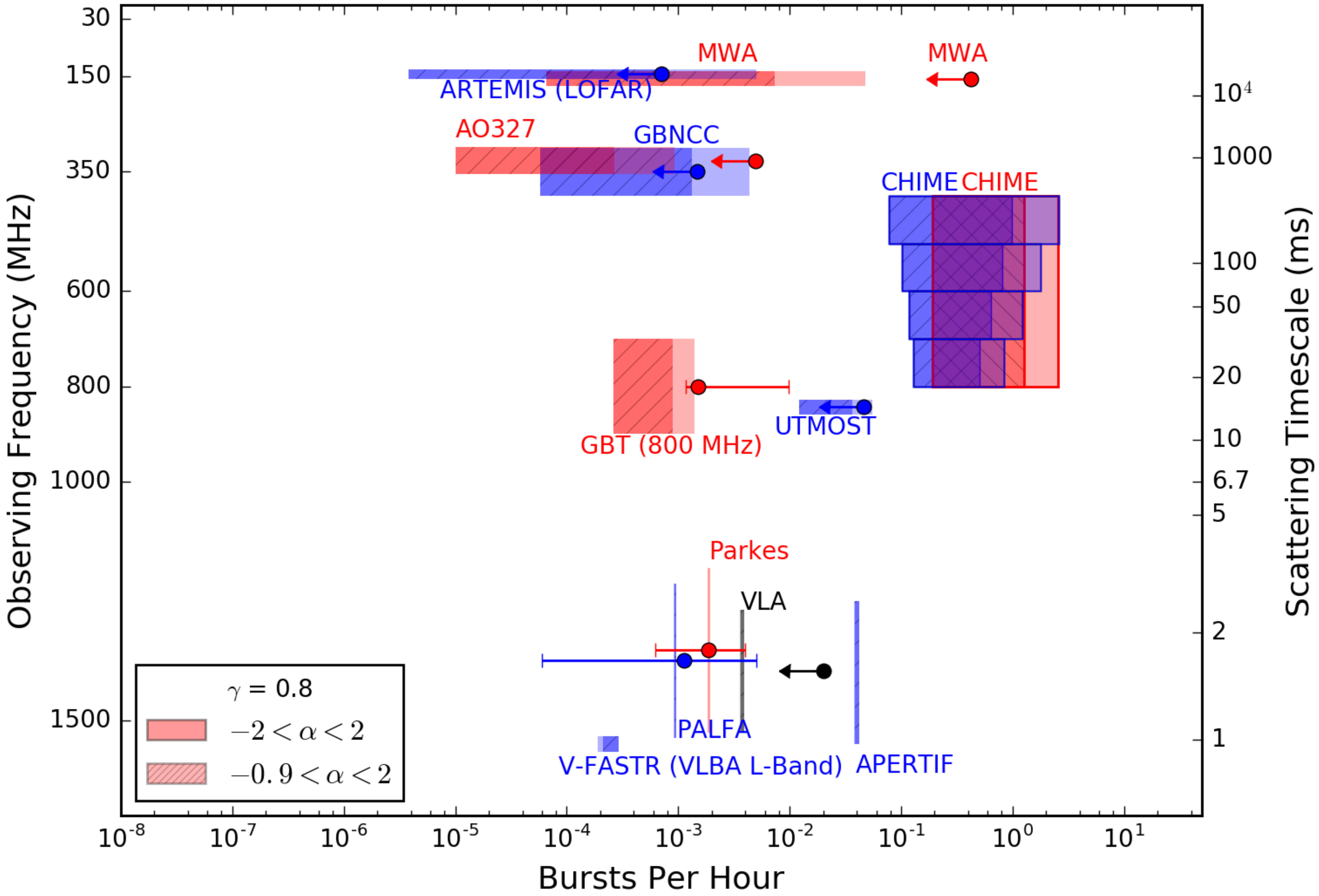}{0.55\textwidth}{(c)}}
\caption{Predicted number of detections per hour for current and upcoming FRB surveys computed by Monte-Carlo simulations using the Parkes rate reported by \citet{crawford2016} as reference. The shaded regions correspond to the bursts per hour for an arbitrary range of spectral indices ($-2 < \alpha <  +2$) while the hatched regions correspond to the bursts per hour for the range, $\alpha_\mathrm{lim} < \alpha <  +2$, where $\alpha_\mathrm{lim}$ for each $\gamma$ is the spectral index constraint derived in this paper. The shaded and hatched regions, as well as the markers denoting the published upper limits on the FRB rate (obtained from the corresponding references in Table \ref{tab:params}) have been colored differently where there are several surveys in a certain frequency range to distinguish between the limits reported by each of these surveys. For MWA, the upper limit reported by \citet{tingay2015} has been plotted here. Panels (a), (b) and (c) show the predicted rates for $\gamma$ = 1.5, 1.2 and 0.8, respectively. The blue shaded and hatched regions in the frequency range of 400--800 MHz represent the varying detection rates for four parts of the CHIME band with a bandwidth of 100 MHz each. Red-colored regions in the same frequency range represent the overall detection rate for CHIME which has a bandwidth of 400 MHz. The regions corresponding to the four parts of the CHIME band are labelled in Panel (a) with the labels denoting the center frequencies of these parts, namely, 450 MHz, 550 MHz, 650 MHz and 750 MHz.}
\label{fig:CHIME}
\end{figure*}

The rate predictions for all surveys that we considered are shown in Figure \ref{fig:CHIME}. Our simulations predict rates that are consistent with the upper limits reported for LOFAR (\citealt{coenen2014}; \citealt{karastergiou2015}), AO327 \citep{deneva2016}, MWA \citep{tingay2015}, VLA \citep{law2015} and UTMOST \citep{caleb2016}. The simulations do not account for repeating sources and the rate reported for the PALFA survey \citep{scholz2016} is based on the detection of a single event. Since our simulations calculate the rate for UTMOST at its full sensitivity, the upper limit shown in Figure \ref{fig:CHIME} is calculated by scaling the reported upper limit for a fluence threshold of 11 Jy ms \citep{caleb2016} to the fluence for the fully sensitive UTMOST survey calculated using Equation \ref{eq:smin}. 

One caveat, however, is that we have difficulty matching the predicted rate for the Parkes surveys with the observations if the mean of the scattering time distribution is set to be the observed 6.7 ms at 1 GHz. This suggests that one or all of the following assumptions: Kolmogorov spectrum, log-normal distribution of scattering timescales, width of the distribution equal to that of the Galaxy might be incorrect. A more sophisticated treatment of the scattering timescale distribution will allow us to make better predictions. 

The results of the simulations also demonstrate the effect of the slope of the log $N$-log $S$ function on the FRB yield of a survey. The log $N$-log $S$ function determines whether field of view or sensitivity is a more important factor for FRB detection. Our simulations predict a greater FRB detection rate for PALFA as compared to Parkes for $\gamma > 1$ but the rates are consistent with each other for $\gamma < 1$. The abundance of fainter bursts implied by $\gamma > 1$ explains the higher rate prediction for PALFA whose greater sensitivity is highly advantageous for FRB detection in that scenario. However, if $\gamma < 1$, there will be an abundance of brighter bursts, thereby allowing the greater field of view of Parkes as compared to PALFA to compensate for the reduction in sensitivity and have a similar FRB detection rate per hour. 

\subsection{Predictions for CHIME}

The simulations suggest that CHIME will be detecting more FRBs than any existing telescope due to its large field of view. However, scattering can reduce the number of detections in the lower part of the band. To model the competing effects of increase in field of view and increase in scattering timescales at lower frequencies, the CHIME bandwidth has been divided into 4 equal parts (centered at 450 MHz, 550 MHz, 650 MHz and 750 MHz) in our simulations. The scattering timescale and field of view for each part has been calculated using its center frequency and the sensitivity has been evaluated using Equation \ref{eq:smin}, assuming a bandwidth of 100 MHz.

For the lower part of the CHIME band (400--500 MHz), our simulations predict 0.5--21 bursts a day for $\gamma = 1.5$ and 2--24 bursts a day for $\gamma = 0.8$. The prediction for the upper part of the CHIME band (700--800 MHz) is 2--15 bursts per day, assuming a Euclidean flux distribution ($\gamma = 1.5$). \citet{connor2016} predict detection of 2--40 bursts per day for the same part of the band based on the one FRB detected with GBT at 800 MHz. Although our rate prediction for 700--800 MHz is not very different from the \citet{connor2016} prediction, there are significant differences in the method of rate estimation. We extrapolate the 1.4-GHz rate estimate reported by \citet{crawford2016} to the frequency in consideration (750 MHz), assuming a distribution of spectral indices and scattering timescales for the FRB population. On the other hand, \citet{connor2016} predict the detection rate based on the measured FRB rate in the relevant frequency range (700--800 MHz) and neglect the distribution of scattering timescales. The overall rate predicted by our simulations, 3--54 bursts per day, is also in agreement with the prediction of detection of 30--100 FRBs per day by \citet{rajwade2016} assuming a cosmological population of FRBs.

\section{Summary \& Conclusions}\label{sec:summary}
We did not detect any FRBs in GBNCC survey pointings amounting to a total observing time of 84 days. The non-detection allows us to determine a 95\% confidence upper limit on the FRB rate at 350 MHz of 3.6$\times 10^3$ FRBs sky$^{-1}$ day$^{-1}$ above a peak flux density of 0.63 Jy for bursts with an intrinsic width of 5 ms. The threshold flux density of the survey ranges from 0.3 Jy for an FRB of 16 ms duration to 9 Jy for a 0.35 ms duration FRB. 

We computed constraints on the mean intrinsic spectral index by performing Monte Carlo simulations of a population of FRBs consistent with the 1.4-GHz rate estimate and assuming a power-law flux density model for FRBs. The FRBs generated in these simulations had spectral indices sampled from a normal distribution and scattering timescales sampled from a log-normal distribution. If intrinsic spectral index were the only reason for our non-detection, i.e. scattering and free-free absorption were absent, the non-detection with GBNCC would be compatible with the Parkes rate estimate reported by \citet{crawford2016} for $\alpha > +0.35$. \citet{karastergiou2015} derived a constraint, $\alpha > +0.1$, based on non-detection with LOFAR at 145 MHz. Non-detection with MWA at 155 MHz implied $\alpha > -1.2$ \citep{tingay2015}. The GBNCC survey, owing to its large observing time and greater sensitivity, thus enables us to place a stronger constraint on spectral index than has any previous survey. 

However, scattering is one possible reason for our non-detection. Another variant of the simulations was aimed at finding the mean scattering timescale that would render FRBs expected for a particular value of spectral index undetectable with GBNCC. Given the observed range of scattering times at 1.4 GHz, we constrain $\alpha > -0.3$ for a Euclidean flux distribution, in the absence of free-free absorption. The constraints on spectral index are very sensitive to the 1.4-GHz rate estimate used in the simulations. The above-mentioned constraint is derived using the \citet{crawford2016} rate estimate. If the rate estimate reported by \citet{champion2016} is used, then the constraint is weaker with $\alpha > -0.9$. The simulations used for deriving these constraints assume a scattering timescale distribution resembling the distribution of Earth-centered scattering times for our Galaxy. However, the scattering timescale depends on the location in, orientation and type of the host galaxy. Detailed treatment of this problem is beyond the scope of this paper. The simulations are also based on the assumption of a power-law spectral model. Although the assumption is in line with previous studies, it could be incorrect if the repeating FRB is a member of the same source class as the rest of the population since observations of the repeating FRB121102 \citep{scholz2016} show that a single power-law is a poor characterization of the burst spectra.

We find that the strongest constraint is obtained for the case of the Euclidean flux distribution, both in the absence of scattering and free-free absorption and in the presence of scattering. A higher value of $\gamma$ corresponds to an increase in the relative abundance of fainter FRBs. Therefore, an increase in $\gamma$ implies an increase in the number of detections with GBNCC by virtue of its sensitivity, thereby requiring higher mean scattering times or a more positive spectral index to explain our non-detection. 

For the particular case of standard candles with a constant comoving number density, we estimate a maximal redshift of 0.37 being probed by the GBNCC survey. We find a spectral index $\alpha_\mathrm{lim}$ for which the peak flux density of an FRB at $z$ = 0.37 is equal to the survey sensitivity. We rejected any spectral index $< \alpha_\mathrm{lim}$ as it would predict sensitivity to a greater redshift and hence detection of FRBs with GBNCC.  In the scenario of free-free absorption with a hot ionized magnetar ejecta, we obtain $\alpha_\mathrm{lim} = -0.6$ and for a cold molecular cloud having ionization fronts, $\alpha_\mathrm{lim} = 1.0$ under the assumption of no scattering. Our constraints imply that spectra of FRBs are different from observed pulsar spectra, for which the mean spectral index is $-$1.4 \citep{bates2013}. However, if FRBs are subject to both free-free absorption and scattering, our constraints are far weaker and allow for steep negative spectral indices as well. 

We also predict the detection rate for existing surveys and upcoming ones such as CHIME using Monte Carlo simulations. The simulations for a Euclidean flux distribution predict that CHIME will detect 3--54 bursts per day assuming the \citet{crawford2016} rate estimate and 1--25 bursts a day assuming the rate estimate reported by \citet{champion2016}. The predictions are promising because even with the most conservative estimates, CHIME will be able to greatly increase the number of known FRBs and probe the distribution of their properties such as spectral index, scattering timescales and the slope of the log $N$-log $S$ function. 

\section{Acknowledgements}
The National Radio Astronomy Observatory is a facility of the National Science Foundation operated under cooperative agreement by Associated Universities, Inc. We thank Compute
Canada and the McGill Center for High Performance Computing and Calcul Quebec for provision and maintenance of the Guillimin supercomputer and related resources. We also thank an anonymous referee for useful comments which helped improve the manuscript. We are grateful to Erik Madsen for providing code to make plots for Figure \ref{fig:CHIME}. PC acknowledges support from a Mitacs Globalink Graduate Fellowship and the TOEFL Scholarship Program in India. VMK receives support from an NSERC Discovery Grant, an Accelerator Supplement and from the Gerhard Herzberg Award, an R. Howard Webster Foundation Fellowship from the Canadian Institute for Advanced Research, the Canada Research Chairs Program, and the Lorne Trottier Chair in Astrophysics and Cosmology. JWTH and VIK acknowledge support from the European Research Council under the European Union's Seventh Framework Programme (FP/2007-2013) / ERC Grant Agreement nr. 337062. MAM is supported by NSF AST Award \#1211701. Pulsar research at UBC is supported by an NSERC Discovery Grant and by the Canadian Institute for Advanced Research. JvL acknowledges funding from the European Research Council under the European Union's Seventh Framework Programme (FP/2007-2013) / ERC Grant Agreement n. 617199. 

\software{PRESTO \citep{ransom2001}, RRATtrap \citep{karako2015}}


\begin{thebibliography}{}
\bibitem[Ade et~al.(2016)]{ade2015} Ade, P.~A.~R., Aghanim, N., Arnaud, M., et al. 2016, A\&A, 594, A13
\bibitem[Anderberg(1973)]{anderberg1973} Anderberg, M.R. 1973, Cluster Analysis for Applications (New York: Academic Press)
\bibitem[Bandura et al.(2014)]{bandura2014} Bandura, K., et. al. 2014, Society of Photo-Optical Instrumentation Engineers (SPIE) Conference Series, Vol. 9145, p. 22
\bibitem[Bannister \& Madsen(2014)]{bannister2014} Bannister, K.~W. \& Madsen, G.~J. 2014, MNRAS, 440, 353
\bibitem[Bates et~al.(2013)]{bates2013} Bates, S.~D., Lorimer, D.~R. \& Verbiest, J.~P.~W., 2013, MNRAS, 431, 1352
\bibitem[Bentley(1975)]{bentley1975} Bentley, J.~L. 1975, Communications of the ACM, 18(9):509-517
\bibitem[Burke-Spolaor \& Bannister(2014)]{BSB2014} Burke-Spolaor~S. \& Bannister K.~W. \ 2014, ApJ, 792, 19 
\bibitem[Burke-Spolaor et~al.(2016)]{spolaor2016} Burke-Spolaor,~S., Trott, C.~M., Brisken, W.~F., et al. 2016, \apj, 826, 2
\bibitem[Caleb et~al.(2016)]{caleb2016} Caleb, M., Flynn, C., Bailes, M., et al. 2016, MNRAS, 458, 718
\bibitem[Champion et~al.(2016)]{champion2016} Champion, D.~J., Petroff, E., Kramer, M., et al. 2016, MNRAS, 460, L30
\bibitem[Chatterjee et~al.(2017)]{chatterjee2017} Chatterjee, S., Law, C.~J., Wharton, R.~S., et al. 2017, Nature, 541, 58 
\bibitem[Coenen et~al.(2014)]{coenen2014} Coenen, T., van Leeuwen, J., Hessels, J. W. T., et al. 2014, A\&A, 570, A60
\bibitem[Connor et~al.(2016)]{connor2016} Connor, L., Lin, H.-H., Masui, K., et al. 2016, MNRAS, 460, 1054
\bibitem[Cordes \& Lazio(2002)]{cordes2002} Cordes, J.~M. \& Lazio, T.~J.~W. 2002, astro-ph/0207156
\bibitem[Cordes \& McLaughlin(2003)]{cordes2003} Cordes, J. M. \& McLaughlin, M.~A. 2003, \apj, 596, 1142
\bibitem[Cordes et~al.(2016)]{cordes2016} Cordes, J.~M., Wharton, R.~S., Spitler, L.~G., et al. 2016, ArXiv e-prints, arXiv:1605.05890
\bibitem[Cordes \& Wasserman(2016)]{cordes2016b} Cordes, J.~M \& Wasserman, I. 2016, MNRAS, 457, 232
\bibitem[Crawford et~al.(2016)]{crawford2016} Crawford, F., Rane, A., Tran, L., et al. 2016, MNRAS, 460, 3370
\bibitem[Deneva et~al.(2016)]{deneva2016} Deneva, J.~S., Stovall, K.,  McLaughlin, M.~A., et al. 2016, \apj, 821, 10
\bibitem[Ester et~al.(1996)]{ester1996} Ester, M., Kriegel, H.-P., Sander, J. \& Xu, X. 1996, Proc. 2nd Int. Conf. on Knowledge Discovery and Data Mining (Portland, OR: AAAI Press), 226
\bibitem[Haslam et~al.(1982)]{haslam1982} Haslam, C.~G.~T, Salter, C.~J., Stoffel, H. \& Wilson, W.~E. 1982, A\&AS, 47, 1
\bibitem[Inoue(2004)]{inoue2004} Inoue, S. 2004, MNRAS, 348, 999
\bibitem[Ioka(2003)]{ioka2003} Ioka, K. 2003, \apj, 598, L79 
\bibitem[Karako-Argaman et~al.(2015)]{karako2015} Karako-Argaman, C., Kaspi, V.~M., Lynch, R.~S., et al. 2015, \apj, 809,67
\bibitem[Karastergiou et~al.(2015)]{karastergiou2015} Karastergiou, A., Chennamangalam, J., Armour, W., et al. 2015, MNRAS, 452, 1254
\bibitem[Katz(2016)]{katz2016} Katz, J.~I. 2016, Mod. Phys. Lett. A, 31, 1630013 
\bibitem[Keane et~al.(2012)]{keane2012} Keane, E.~F., Stappers, B.~W., Kramer, M., \& Lyne, A. G. 2012, MNRAS, 425, L71
\bibitem[Keane et~al.(2016)]{keane2016} Keane, E.~F., Johnston, S., Bhandari, S., et al. 2016, Nature, 530, 453
\bibitem[Kulkarni et~al.(2015)]{kulkarni2015} Kulkarni, S.~R., Ofek, E.~O., \& Neill, J.~D. 2015, ArXiv e-prints, arXiv:1511.09137
\bibitem[Law et~al.(2015)]{law2015} Law, C.~J., Bower, G.~C., Burke-Spolaor, S., et al. 2015, \apj, 807, 16
\bibitem[Lorimer \& Kramer(2005)]{lorimer2005} Lorimer, D.~R. and Kramer, M. 2005, Handbook of Pulsar Astronomy (Cambridge University Press)
\bibitem[Lorimer et~al.(2007)]{lorimer2007} Lorimer, D.~R., Bailes, M., McLaughlin, M.~A., et al. 2007, Science, 318, 777
\bibitem[Lorimer et~al.(2013)]{lorimer2013} Lorimer, D.~R., Karastergiou, A., McLaughlin, M.~A. \& Johnston, S. 2013, MNRAS, 436, L5
\bibitem[Macquart \& Johnston(2015)]{macquart2015} Macquart, J.-P., \& Johnston, S. 2015, MNRAS, 451, 3278
\bibitem[Masui et~al.(2015)]{masui2015} Masui, K., Lin, H.-H.,  Seivers, J., et al. 2015, Nature, 528, 523 
\bibitem[Mezger \& Henderson(1967)]{mezger1967} Mezger, P.~G. \& Henderson, A.~P. 1967, \apj, 147, 471
\bibitem[Oppermann et~al.(2016)]{oppermann2016} Oppermann, N., Connor, L.~D. \& Pen, U. 2016, MNRAS, 461, 984
\bibitem[Petroff et~al.(2014)]{petroff2014} Petroff, E., van Straten, W., Johnston, S. 2014, ApJL, 789, 2
\bibitem[Petroff et~al.(2015)]{petroff2015} Petroff, E., Bailes, M., Barr, E.~D., et al. 2015, MNRAS, 447, 246
\bibitem[Petroff et~al.(2016)]{petroff2016} Petroff, E., Barr, E.~D., Jameson, A., et al. 2016, Publications of the Astronomical Society of Australia, 33, 45
\bibitem[Popov \& Postnov(2013)]{popov2013} Popov, S.~B. \& Postnov, K.~A. 2013, ArXiv e-prints, arXiv:1307.4924 
\bibitem[Rajwade \& Lorimer(2017)]{rajwade2016} Rajwade, K.~M. \& Lorimer, D.~R., 2017, MNRAS, 465, 2
\bibitem[Ransom(2001)]{ransom2001} Ransom, S.~M. 2001, PhD thesis, Harvard University
\bibitem[Ravi et~al.(2015)]{ravi2015} Ravi, V., Shannon, R.~M., \& Jameson, A. 2015, ApJL, 799, L5
\bibitem[Ravi et~al.(2016)]{ravi2016} Ravi, V. \& Shannon, R.~M. et al., 2016, Science
\bibitem[Remazeilles et~al.(2015)] {remazeilles2015} Remazeilles, M., Dickinson, C., Banday, A.~J., et al. 2015, MNRAS, 451, 4311
\bibitem[Rowlinson et~al.(2016)]{rowlinson2016} Rowlinson, A., Bell, M.~E., Murphy, T., et al. 2016, MNRAS, 458, 3506
\bibitem[Scholz et~al.(2016)]{scholz2016} Scholz, P., Spitler, L.~G., Hessels, J.~W.~T., et al. 2016, \apj, 833, 2
\bibitem[Spitler et~al.(2014)]{spitler2014} Spitler, L.~G., Cordes, J.~M., Hessels, J.~W.~T., et al. 2014, \apj, 790, 101 
\bibitem[Spitler et~al.(2016)]{spitler2016} Spitler, L.~G., Scholz, P., Hessels, J.~W.~T., et al. 2016, Nature, 531, 202
\bibitem[Stovall et~al.(2014)]{stovall2014} Stovall, K., Lynch, R.~S., Ransom, S.~M., et al. 2014, \apj, 791, 67
\bibitem[Thornton et~al.(2013)]{thornton2013} Thornton, D., Stappers, B., Bailes, M., et al. 2013, Science, 341, 53
\bibitem[Tendulkar et~al.(2017)]{tendulkar2017} Tendulkar, S.~P, Bassa, C.~G., Cordes, J.~M. et al. 2017, ApJL, 834, 2
\bibitem[Tingay et~al.(2013)]{tingay2013} Tingay, S.~J., Goeke, R., Bowman, J.~D., et al. 2013, PASA, 30, 7
\bibitem[Tingay et~al.(2015)]{tingay2015} Tingay, S.~J., Trott, C.~M., Wayth, R.~B., et al. 2015, \apj, 150, 6 
\bibitem[van Leeuwen(2014)]{vanleeuwen2014} van Leeuwen, J. 2014, in The Third Hot-wiring the Transient Universe Workshop, ed. P. R. Wozniak, M.J. Graham, A. A. Mahabal, \& R. Seaman, 79, http://www.slac.stanford.edu/econf/C131113.1/
\bibitem[Vander Wiel et~al.(2016)]{vanderwiel2016} Vander Wiel, S., Burke-Spolaor, S., Lawrence, E., et al. 2016, ArXiv e-prints, arXiv:1612.00896 
\bibitem[Vedantham et~al.(2016a)]{vedantham2016a} Vedantham, H.~K., Ravi, V., Mooley, K., et al. 2016a, ApJL, 824, L9
\bibitem[Vedantham et~al.(2016b)]{vedantham2016b} Vedantham, H.~K., Ravi, V., Hallinan, G., et al. 2016b, \apj, 830, 75 
\bibitem[Wayth et~al.(2012)]{wayth2012} Wayth, R.~B., Tingay, S.~J., Deller, A.~T., et al. 2012, ApJL, 753, 2
\bibitem[Williams \& Berger(2016)]{williams2016} Williams, P.~K.~G., \& Berger, E. 2016, ApJL, 821, L22

\end{thebibliography}
\end{document}